# Solution to an unsolved problem in diodes: Limiting current for small emission area and low emission energy

**Y. Y. Lau[1], David P. Chernin[2], Yves Heri[1], and Peng Zhang[1]**

[1]Department of Nuclear Engineering and Radiological Sciences, University of Michigan, Ann Arbor, MI 48109, USA

[2]Leidos, Inc., Reston, VA 20190, USA

## ABSTRACT

The maximum current that can be extracted from a diode is a central question in electronic devices, especially for the generation of radiation from microwaves to x-rays. The challenge in its prediction increases significantly when the electron emission is restricted to a small area for which the classical one-dimensional Child-Langmuir law is no longer applicable. We address this unsolved problem using a model that simultaneously includes a small emission area and a small electron emission velocity. New scaling laws are presented for this difficult regime. These results are obtained from three vastly different approaches: a differential equation formulation which provides extremely high resolution, an integral equation formulation which leads to the scaling laws, and a particle-in-cell simulation which shows the temporal-spatial evolution. Comparisons of the predicted maximum current among these three approaches are performed over a large range of parameters, paying special attention to the resolution of the potential minimum in the immediate vicinity of the cathode surface. Corroborations of the scaling laws with experiments on thermionic cathodes and on photoinjectors are indicated. The model consists of a periodic array of electron sheets of finite width in a planar diode, all emitted from a cathode with the same energy. Electron motion is restricted to the direction normal to the cathode. This problem is of interest to the electron device, accelerator, pulsed power, aerospace, and applied mathematics communities.

## I. INTRODUCTION

High intensity electron beams are used to power modern radiation sources ranging from x-ray [1], ultraviolet [2], and terahertz, down to millimeter wave and microwave [3]. The brightness of an electron beam is a critical limiting factor for the brightness of the radiation that the electron beam is used to generate. The electron beam brightness, in turn, is limited by the maximum current that may be extracted from a diode through an anode [4][5][6]. To lowest order, this maximum current is approximately governed by the classical, one-dimensional (1D) Child-Langmuir Law (CLL) [7][8], which gives the maximum, uniform steady state current density ($J_{CL}$) that can be transported across a planar diode. Because of the prevalence of electron emission from local regions, and the inevitable non-uniform electron emission from a cathode surface, there have been numerous attempts to extend the 1D CLL to 2D and 3D computationally [9]-[27], phenomenologically [28]-[34], and analytically [35]-[44] for various applications. None succeeded in the notoriously difficult regime of simultaneous small emission site and small initial electron emission velocity. This paper addresses this problem. Specifically, we report analytical and numerical results from three very different approaches, including derivations of new scaling laws for the limiting current in this complex regime.

This classical 2D limiting current problem is of current interest to the accelerator, pulsed power, electron device, and aerospace communities. Despite its classical nature, we note that the various quantum mechanical emission mechanisms (e.g., thermionic, field, and photoemission) reach physical limits in real diodes due to classical space-charge effects that are well-captured by the 1D CLL and its extensions to higher dimensions.

The 1D CLL for a planar diode is [7][8],

$$J_{CL} = \frac{4\sqrt{2}}{9}\epsilon_0\sqrt{\frac{e}{m}}\frac{V^{3/2}}{d^2}\,, \tag{1}$$

in SI units, where $V$ is the diode voltage, $d$ is the gap separation of the diode, $e$ and $m$ are respectively the electron charge and mass, and $\epsilon_0$ is the free space permittivity. Electrons are assumed to be emitted with zero initial velocity. Extensions of Eq. (1) to 2D have proven to be extremely difficult because of the intricate interplay of geometry, dimensionality, and injection conditions. By fitting results on limiting current obtained from particle-in-cell (PIC) simulations [10], and supported analytically [35], the 2D CLL for an infinitely long emission stripe of a finite width $w$, and for a circular emission patch of radius $R$, were found to be, respectively,

$$\frac{J_{CL-2D}}{J_{CL}} \cong 1 + \frac{d}{\pi w}\,, \qquad 0 < d/w < 10, \tag{2a}$$

$$\frac{J_{CL-2D}}{J_{CL}} \cong 1 + \frac{d}{4R}\,, \qquad 0 < d/R < 2, \tag{2b}$$

assuming uniform emission current density over the 2D patch on the cathode of a planar diode. Note that the scaling laws (2a) and (2b) are valid only for sufficiently large $w$ and $R$, i.e., only for sizeable emission regions compared with the gap size $d$. If there is a sufficient supply of electrons in the emission patch to drive the electric field to zero everywhere on that patch (also known as the space-charge-limited condition), the uniform emission model is inapplicable and PIC simulations [11][12][37] revealed that the emitted current density profile exhibits a wing-like structure at the edges of the emitting patch, where the local current density could also be much higher than the 1D CLL value. These 2D effects are particularly pronounced for small emitting patches, a trend also suggested by Eqs. (2a) and (2b). They lead to the recent discovery [37][38] that the anode current in a realistic, highly nonuniformly emitting thermionic cathode predominantly originates from the regions of the lowest work function at high temperatures, even though such regions occupy only a very small fraction of the total cathode area. Taking this trend to the extreme, it is shown that for emission sheets of vanishing width, the local current density can exceed Eq. (1) by an arbitrarily large factor [39][40]. Nonetheless, these papers conclude that the total anode current, averaged over the entire cathode surface with non-uniform emission, is bounded by but could approach the 1D CLL, a remarkable finding. If all electrons are injected into the diode with a constant, unidirectional initial velocity $v_{in}$, Eq. (1) is enhanced by the Jaffe factor [45], $f_J = \left[(1+\Delta)^{1/2} + \Delta^{1/2}\right]^3$, where $\Delta = E_{in}/eV$ and $E_{in} = mv_{in}^2/2$. Virtual cathodes are formed when the critical current is reached [46][47]. Kinetic theory, direct solutions of the Poisson–Vlasov system, and PIC simulations have been used to study virtual cathode phenomena, which are inevitable with an excessive supply of electrons [46]-[50]. The transitions between thermionic, field-emission, and space-charge-limited regime are studied in [6], [51]-[59].

Despite the extensive body of work, a critical omission remains in the simultaneous limit of small emission area and small initial velocity of electrons. Existing multidimensional theories typically rely on the assumptions – such as sizeable emission area, finite injection energy, or quasi-one-dimensional transport that break down in this regime. Numerical approaches face resolution and scaling challenges due to the coexistence of disparate spatial and velocity scale. As a result, a predictive and unified scaling law that captures the coupled effects of emitter size, gap spacing, and injection velocity has been lacking. Insight into this regime is much desired for the basic understanding and for the design of nanoscale emitters [58][60][52], advanced vacuum electronic devices [6][50][60], photo injectors [19]-[21], extreme ultraviolet and x-ray sources [1][2][61], and even the transpiration cooling in hypersonic aircraft via electron emission [62][63], to name just a few applications of contemporary interest. Here, we address this outstanding problem by combining three complementary approaches: differential equation, integral equation, and PIC simulation.

The model under study is an infinite periodic array of finite width sheet electron beams in a planar diode, a single period of which is illustrated in Fig. 1. The diode has gap separation $d$, gap voltage $V$, the width of the electron sheet is $w$, and the period of the array is $p$. We assume that the motion of the emitted electrons is restricted by an arbitrarily large magnetic field to the $z$-direction and that all electrons are emitted at $z = 0$ with the same kinetic energy $E_{in}$. The emission current density $J$ is constant within each electron sheet and is zero elsewhere. All quantities are independent of $y$. We seek to determine the maximum value of $J \equiv J_{max}$, for given values of $d$, $V$, $w$, $p$, and $E_{in}$. Upon normalization, we find $J_{max}/J_{CL}$ dependent on the three dimensionless parameters, $\bar{w} = w/d$, $\Delta = E_{in}/eV$, and $\bar{p} = p/d$. These three parameters measure, respectively, the emitter width, the electron emission energy, and the packing density of the periodic array. Thus, our results are quite general; in particular, they are independent of the electron emission mechanism and composition in the cathode material, just like the Child-Langmuir law.

Through a systematic study of the variation of the three parameters $(\bar{w}, \Delta, \bar{p})$, we compare the numerical results obtained from the three (3) different approaches, the differential equation approach (results from which are designated POISSON, below), the integral equation approach, and particle-in-cell simulation (designated XOOPIC [64]). POISSON and XOOPIC are the names of computer codes we applied to this problem. We concentrate on the difficult cases with $(\bar{w}, \Delta) \rightarrow (0,0)$ and $\bar{p} \gg 1$. The last condition implies that each electron sheet is essentially isolated from its neighbors, and our extensive numerical data, reported below, suggest that $\bar{p} > 2$ suffices. Results including parameters outside this range of $\bar{p}$ will also be presented to show the effects of neighboring electron sheets on the maximum anode current.

The differential equation approach (POISSON) offers the highest resolution. To illustrate, consider the extremely demanding case with $(\bar{w}, \Delta, \bar{p}) = (10^{-3}, 10^{-5}, 2)$, where the potential minimum is found to occur at a distance of $4 \times 10^{-5}d$ from the cathode surface, and the value of $J_{max}/J_{CL}$ is to be determined to within an accuracy of 1 part in 1000. Such an ultra-high resolution result from POISSON, using 100,000 cells across the gap and 16,384 cells per period (Fig. 1), perhaps for the first time, allows a definitive quantification of the uncertainties in the numerical determination of the limiting current using PIC (XOOPIC) whose uncertainties are most acute when $(\bar{w}, \Delta) \rightarrow (0,0)$. We discover from the integral equation approach that there are two very different scaling laws for $J_{max}/J_{CL}$ as $(\bar{w}, \Delta) \rightarrow (0,0)$, depending on whether $2\pi\Delta \ll \bar{w}$ or $2\pi\Delta \gg \bar{w}$. Such a difference is traced analytically to the relative sizes of the two terms, $\partial E_x/\partial x$ and $\partial E_z/\partial z$, in the Poisson equation near the potential minimum, and this observation is confirmed by the high-resolution results of POISSON. In their respective regimes of validity, these scaling laws for $J_{max}/J_{CL}$ obtained from the integral equation compare well with the numerical results obtained from POISSON and XOOPIC, and the scaling laws are corroborated in experiments on thermionic cathodes and photoinjectors.

In Section II, we present the major results and summarize the novel findings. In Section III, we outline the three independent methods used: differential equation, integral equation, and PIC simulation. In Section IV, we offer additional insight and discuss various issues. The details of the derivation of the scaling laws, and the high-resolution numerical data, are given in the Appendixes.

## II. RESULTS

The major findings are summarized at the end of this section, after presenting the numerical data obtained from the three approaches: POISSON, XOOPIC, and the scaling laws obtained from the integral equation.

Among the three dimensionless parameters, $(\bar{w}, \Delta, \bar{p})$, we first systematically vary $(\bar{w}, \Delta)$, but set $\bar{p} \geq 2$ so that each electron sheet is essentially isolated from its neighbors, a result that we shall demonstrate when we include the $\bar{p} < 1$ cases. Three groups of tests were performed: (A) The first group consists of three blind tests (E1-E3) for validation, in which $\bar{w}$ and $\Delta$ assume small, but moderate values $(\bar{w} \geq 0.01, \Delta \geq 0.001)$. (B) The second group consists of three new tests, E4-E6, which together with E3, provide four tests on $\Delta$, $\Delta = 10^{-2}, 10^{-3}, 10^{-4}, 10^{-5}$. We fix $\bar{w} = 0.01$ for this group. (C) The third group, tests E7-E10, is the most demanding computationally. It is the same as the second group except $\bar{w} = 0.001$. Salient features in the spatial profiles of representative runs are displayed and interpreted. *We shall use the POISSON code results as the standard* because POISSON offers the highest resolution for small, but finite values of $(\bar{w}, \Delta)$. In all POISSON and XOOPIC simulations, we fix $V = 2.5$ kV, $d = 1$ cm and vary $(\bar{w}, \Delta, \bar{p})$. In Appendix D, we list the parameters used, and the numerical results obtained from POISSON, for the various tests reported in this paper, in particular Tests E1-E10, and other tests at smaller values of $\bar{p}$.

Figure 2 shows the comparison of $J_{max}/J_{CL}$ of Tests E1-E10 using POISSON, XOOPIC, and the scaling laws, Eqs. (3) and (4). These scaling laws are valid for the regimes of small $\bar{w}$ and $\Delta$, depending on whether $2\pi\Delta \ll \bar{w}$ [Eq. (3)], or $\bar{w} \ll 2\pi\Delta$ [Eq. (4)],

$$\frac{J_{max}}{J_{CL}} \cong \frac{9}{8} \times \frac{1}{\sqrt{\bar{w}}}, \qquad 2\pi\Delta \ll \bar{w} < 0.1, \tag{3}$$

$$\frac{J_{max}}{J_{CL}} \cong (3.302/\bar{w})\Delta^{0.5274}, \qquad \bar{w}/2\pi \ll \Delta < 0.01. \tag{4}$$

These scaling laws agree well with the POISSON and XOOPIC codes in their respective regimes, as shown in Fig. 2. Section III outlines the method used to derive Eqs. (3) and (4), which are valid for $\bar{p} \geq 2$, with the details given in Appendices B and C. These Appendices further show that when $2\pi\Delta \ll \bar{w}$, the term $\partial E_z/\partial z$ is dominant and that when $\bar{w} \ll 2\pi\Delta$, the term $\partial E_x/\partial x$ is dominant in the Poisson equation near the potential minimum as $J \to J_{max}$. Figure 3 shows that this is indeed the case for Tests E7 and E10, where we present the very high-resolution spatial distributions of the potential $\phi$, the charge density $\rho$, and the electric field components $E_z$ and $E_x$. In Fig. 3 and elsewhere, $n_z$ is the number of numerical cells across the diode, $n_x$ is the number of cells from $x = -p/2$ to $x = +p/2$; $d_z = d/n_z$ and $d_x = p/n_x$ are the respective cell size (Fig. 1).

Figure 4 shows the spatial-temporal evolution for Test E3 $[(\bar{w}, \Delta, \bar{p}) = (0.01, 0.01, 2)]$ according to XOOPIC, when $J$ is slightly less than and slightly greater than $J_{crit}$, which denotes the value of $J_{max}$ determined from XOOPIC [64]. Both high resolution ($n_z = 1000$) and low resolution ($n_z = 100$) runs are presented to illustrate the dependence of grid size in PIC in this study. The high resolution XOOPIC run gives $J_{crit} = 32.75 J_{CL}$ whereas the POISSON results give $J_{max} = 32.863 J_{CL}$ for Test E3 (see Appendix D). In Fig. 4, we present in (a) the center line electric field on the cathode at which onset of turbulent behavior is used as the criterion to determine $J_{crit}$ [10], (b) the spatial distribution of the electrostatic potential and the potential minimum, (c) and (d) the temporal evolution of the position and the value of the potential minimum, respectively, for $J < J_{crit}$ and $J > J_{crit}$. These figures show the abrupt transition, spatially and temporally, at $J = J_{crit}$, a phenomenon previously known for the model of mono-energetic electron emission velocity [39][47][48][49]. This test case of E3 exemplifies the spatial-temporal behaviors at transition because XOOPIC, POISSON, and the scaling law all yield almost identical results for $J_{max}$, as is apparent in Fig. 2(a).

Figures 5 and 6 include the data for $\bar{p} = p/d < 1$, obtained from POISSON. Figure 5 plots $< \bar{J}_{max} > \equiv (\bar{w}/\bar{p}) J_{max}/(J_{CL} f_J)$, which is the normalized current density including the Jaffe factor $f_J$, averaged over one period of the periodic array (Fig. 1). It shows $< \bar{J}_{max} >$ as a function of $\bar{p}$ at various values of $\Delta$ for the regime, $\bar{w} \ll 2\pi\Delta$. The eight (8) crosses in Fig. 5 show the POISSON data for $\bar{w} = 0.001$, $\Delta = 0.01$ and 0.001, at various values of $p/d$: 0.1, 0.5, 2.038, 3.277 (see Appendix D for the listing of the parameters used for these eight crosses). The triangles in Fig. 5 are obtained from the integral equation formulation in the limit of infinitesimally thin electron sheets, $\bar{w} \to 0$ [40], thus the condition, $\bar{w} \ll 2\pi\Delta$, is automatically satisfied. The solid curves in Fig. 5 are obtained from the scaling law, Eq. (4). Note the excellent agreement between the scaling law (4), the POISSON data, and the integral equation approach for $\bar{p} > 0.5$. A similar investigation (not shown) also demonstrates excellent agreement between the POISSON data and the scaling law (3) for the opposite regime, $\bar{w} \gg 2\pi\Delta$, for $\bar{p} > 0.5$. *This is a surprising result because it says that if the electron sheets are separated from each other by more than* $1/2$ *of the AK gap spacing, each electron sheet may be regarded as isolated from all of its neighbors as far as current transport to the anode is concerned*, as both scaling laws (3) and (4) are derived under the assumption of $\bar{p} \to \infty$ from the integral equation approach. It also says that, for $\bar{p} > 0.5$, one must apply the 2D Child-Langmuir law. [The black solid curve in Fig. 5 shows more deviation from the integral equation approach (black triangles) because $\Delta = 0.1$ on this black solid curve, which is beyond its range of validity, $\Delta < 0.01$, for the scaling law Eq. (4)]. On the other hand, if $\bar{p} \leq 0.1$, the example of Fig. 6 suggests that the average current density may be approximated by the 1D classical Child-Langmuir value, modified by the Jaffe factor for nonzero emission velocity. These novel properties of 1D vs 2D depending on the packing density ($\bar{p}$) of a periodic array have important implications on the design of 2D microelectronic devices. They are also relevant to the shielding effects on field emitter arrays [18]. The parameters used in the POISSON code results of Fig. 6 are listed in Appendix D.

Figure 6 shows that the spatial profiles ($\phi$, $\rho$, $E_x$, $E_z$) look similar near the potential minimum at $J = J_{max}$ for this regime, $\bar{w} \ll 2\pi\Delta$, regardless of the values of $\bar{p}$, whether $\bar{p} \ll 1$ (in which case 1D CLL largely applies) or $\bar{p} \gg 1$ (in which case one must use 2D CLL). Furthermore, the potential and density profiles as a function of $z$ as given in Figs. 6a, and 6c qualitatively look very similar to Figs. 3a, 3c, even though all of these figures encompass the different scalings, Eqs. (3) and (4). Thus, resolution in $z$ alone, which has been the emphasis in most literature on virtual cathodes, cannot reveal the two different scalings (3) and (4) reported in this paper.

To further illustrate the importance of $x$-dependence, we display in Fig. 7 the location ($z_{min}$) and the depth ($\phi_{min}$) of the local potential minimum as a function of $x$ for the two Tests E2 and E3. The marked differences between Figs. 7(a) and 7(b), respectfully for the regimes $2\pi\Delta \ll \bar{w}$ and $\bar{w} \ll 2\pi\Delta$, are noted. Despite such a marked difference in the $x$-dependence for these two different regimes, the $z$-dependence of the potential distribution looks qualitatively similar, as may be seen by comparing Figs. 3a and 3c, which show qualitatively similar $z$-dependence while belonging to these two different regimes. The distribution of the potential minimum for each value of $x$ was rarely explored in previous 2D CL law studies to our knowledge and is potentially important to thermal management and edge cooling. The recognition of the importance in the relative size between $\partial E_z/\partial z$ and $\partial E_x/\partial x$ could be utilized in future practical designs of high-power diodes to control edge emission by field-shaping.

## Summary of the major findings

**A.** For an isolated electron sheet ($\bar{p} = p/d \to \infty$, Fig. 1), the integral equation leads to two scaling laws for $J_{max}/J_{CL}$, Eqs. (3) and (4), depending on whether $2\pi\Delta \ll \bar{w}$ or $\bar{w} \ll 2\pi\Delta$ for small values of $\bar{w}$ and $\Delta$. The discovery of these two separate scalings requires careful quantification of the potential minimum both in $x$- and $z$-directions, as provided in the present paper both analytically and computationally for the first time. They have also been identified with the dominance of the individual term, $\partial E_z/\partial z$ or $\partial E_x/\partial x$, in the Poisson equation at the potential minimum as $J_{max}$ is approached. The high-resolution POISSON data corroborate these statements, as shown in Figs. 2 and 3.

**B.** In the limit of small $\bar{w}$ and $\Delta$, the limiting current density $J_{max}$ for the periodic array (Fig. 1) is largely governed by the 2D Child-Langmuir law for an isolated electron sheet if $\bar{p} > 0.5$ (i.e., as if $\bar{p} \to \infty$). For $\bar{p} < 0.1$, the average current density per period is largely governed by the classical 1D CLL (i.e., as if $\bar{p} \to 0$). Figure 5 shows this transition between 1D to 2D behavior in a periodic array, at the same time displaying the effects of packing density on the maximum anode current. It also demonstrates good agreement between the different approaches.

**C.** We have made significant extensions to the scaling law, Eq. (3). For an isolated circular emitting region of radius $R \ll d$, analogous to Eq. (3), we obtain [cf. Eqs. (B10) and (B11) of Appendix B]

$$\frac{J_{max}}{J_{CL}} \cong \frac{1.328}{\sqrt{R/d}}, \qquad \Delta \ll R/d \ll 1, \tag{5}$$

and the maximum current, $I_{max} \cong 9.72 \times 10^{-6} A \times [V(volt)]^{3/2}(R/d)^{3/2}$. This anode current $I_{max}$ for a circular emission patch has been known as the "cigar regime" in the accelerator community [19][20][21]. We make a further extension and propose a 3-dimensional scaling law for an isolated elliptical emitting region of major radius $R$ and minor radius $w/2$ ($\leq R$),

$$\frac{J_{max}}{J_{CL}} \cong \frac{9}{8}\frac{1}{\sqrt{w/d}} + \frac{0.5325}{\sqrt{R/d}}, \qquad \pi\Delta \ll w/2d \leq R/d \ll 1, \tag{6}$$

so that scaling law (6) reduces to the scaling law (3) in the limit of an infinite stripe ($R \gg w/2$), and to the scaling law (5) in the limit of a circular emitter ($w = 2R$).

**D.** The scaling laws corroborate with experiments. The scaling law (5), which is an extension of the scaling law (3), is recently experimentally verified in photoinjectors by Denham, et al. [20]. The scaling law (4) is found to be consistent with the performance of thermionic cathodes [39]. Thus, this paper fills an important gap with new physical, analytical, and computational insights with the scaling laws, Eq. (3) and Eq. (4), and their extensions, Eqs. (5) and (6).

# III. METHODS

This section outlines the three (3) independent methods used to obtain our solutions for the maximum current density: the differential equation approach whose numerical results are designated POISSON, the integral equation approach that yields the scaling laws (3) and (4), and the PIC simulation whose numerical results are designated XOOPIC. The current density $J$ ($> 0$) is a constant in the region $R$, defined as $-w/2 \leq x \leq w/2$, $0 \leq z \leq d$, and zero elsewhere for the time-independent solutions (Fig. 1).

**Differential Equation Approach**

The charge density is $-J/v(x,z)$, where $v(x,z)$ is the local beam velocity which may be expressed in terms of the potential $\phi(x,z)$ by conservation of energy. Using the normalized variables, $\bar{\phi} = \phi/V,\ \bar{x} = x/d,\ \bar{z} = z/d,\ \bar{p} = p/d, \bar{w} = w/d,$ and $\Delta = E_{in}/\text{eV}$, Poisson's equation for $\phi(x,z)$ becomes,

$$\left(\frac{\partial^2}{\partial \bar{x}^2} + \frac{\partial^2}{\partial \bar{z}^2}\right)\bar{\phi}(\bar{x},\bar{z}) = \begin{cases} \frac{4}{9} f_J \bar{J}[\bar{\phi}(\bar{x},\bar{z}) + \Delta]^{-1/2}, & \text{in } R \\ 0, & \text{elsewhere} \end{cases} \equiv \bar{S}(\bar{x},\bar{z}) \tag{7}$$

where we have defined the normalized current density $\bar{J} \equiv J/(f_J J_{CL})$, and $f_J = \left[(1+\Delta)^{1/2} + \Delta^{1/2}\right]^3$ is the Jaffe factor [45] for non-zero emission energy $E_{in}$. The potential $\bar{\phi}(\bar{x},\bar{z})$ satisfies the boundary conditions $\bar{\phi}(\bar{x},0) = 0, \bar{\phi}(\bar{x},1) = 1$ and $\bar{\phi}(-\bar{p}/2,\bar{z}) = \phi(+\bar{p}/2,\bar{z})$. Our goal is to find the maximum value of $\bar{J} \equiv \bar{J}_{max}(\bar{w},\Delta,\bar{p})$ such that Eq. (7) has a physical (finite, real valued) solution. We note that $\bar{J}_{max} \to 1$ as $\bar{w} \to \infty$, and expect that $\bar{J}_{max} > 1$ for finite values of $\bar{w}$.

We construct an iterative solution of Eq. (7) as follows: Denoting the k-th iterative approximation of the potential as $\phi^{(k)}(x,z)$, we compute the $(k+1)$ iterative approximation as the solution of

$$\left(\frac{\partial^2}{\partial \bar{x}^2} + \frac{\partial^2}{\partial \bar{z}^2}\right)\bar{\phi}^{(k+1)}(\bar{x},\bar{z}) = \bar{S}^{(k)}(\bar{x},\bar{z}). \tag{8}$$

Equation (8) may be solved for $\bar{\phi}^{(k+1)}$ on a discrete 2D grid in $x$ and $z$ using the method described in Appendix B of Chernin et al. [37], by employing a Fourier cosine series expansion in $\bar{x}$ and a tri-diagonal solver in $\bar{z}$. The iteration is initialized with the vacuum solution, $\bar{\phi}^{(0)}(\bar{x},\bar{z}) = \bar{z}$, and continues until one of the following three things happens: (a) the fractional difference between $\bar{\phi}^{(k)}(\bar{x},\bar{z})$ and $\bar{\phi}^{(k-1)}(\bar{x},\bar{z})$ is less than a specified tolerance for all points on the grid, (b) the argument of the square root in the definition of $\bar{S}^{(k)}(\bar{x},\bar{z})$ becomes negative anywhere on the grid, or (c) a maximum number of iterations is reached. Condition (a) defines convergence. Condition (b) or (c) defines non-convergence. We have used a tolerance of $10^{-5}$ in condition (a) and a maximum number of iterations of 300 in condition (c) in the POISSON results presented in this paper.

The value of $\bar{J}_{max}(\bar{w},\Delta,\bar{p})$ is determined using a simple bisection algorithm [40]: For a fixed value of the period $\bar{p} > \bar{w}$ and an assigned pair of nonzero values for $(\bar{w},\Delta)$, we start with a pair of values of $\bar{J}$ that we estimate will bracket the expected value of $\bar{J}_{max}$; in particular, we start with $\bar{J}_1$ and $\bar{J}_2$ such that we anticipate that the iteration (8) will converge for $\bar{J} = \bar{J}_1$ and will not converge for $\bar{J} = \bar{J}_2$. We then try a value of $\bar{J}_{mid} \equiv (\bar{J}_1 + \bar{J}_2)/2$. If the iteration converges, we assign a new value for $\bar{J}_1 = \bar{J}_{mid}$; if the iteration does not converge, we assign a new value for $\bar{J}_2 = \bar{J}_{mid}$. We repeat this bisection 12 times, which typically gives $\bar{J}_{max}$ to an accuracy of ~3-4 digits. For general values of $(\bar{w},\Delta,\bar{p})$, this algorithm yields the fastest and the most accurate value of $J_{max}/J_{CL}$ compared with the integral equation approach and PIC simulation.

**Integral Equation Approach**

For an *isolated* emission stripe with uniform emission current density $J$ over a finite width $w$ (Fig. 1, with $p/d \to \infty$), the electrostatic potential $\phi(x,z)$ is governed by the integral equation,

$$\bar{\phi}(\bar{x},\bar{z}) = \bar{z} + \frac{2}{9\pi}\left(\frac{J}{J_{CL}}\right)\int_0^1 d\bar{z}_c \int_{-\bar{w}/2}^{\bar{w}/2} d\bar{x}_c \frac{1}{(\bar{\phi}(\bar{x}_c,\bar{z}_c) + \Delta)^{1/2}} \bar{G}(\bar{x},\bar{z};\bar{x}_c,\bar{z}_c)\,, \quad 0 \le \bar{z} \le 1, -\infty < \bar{x} < \infty. \tag{9}$$

Equation (9) is derived in Appendix A. In the RHS of Eq. (9), the first term $(\bar{z})$ is the vacuum potential. The second term is the space charge potential, proportional to the current density $J$ and is given in terms of the Green's function $\bar{G}$, which is the potential at $(\bar{x},\bar{z})$ due to a line charge of unit line charge density located at $(\bar{x}_c,\bar{z}_c)$ between two grounded plates at $\bar{z}=0$ and $\bar{z}=1$ [cf. Eq. (A3) in Appendix A],

$$\bar{G}(\bar{x},\bar{z};\bar{x}_c,\bar{z}_c)=\frac{1}{2}\ln\left[\frac{\cosh\pi(\bar{x}-\bar{x}_c)-\cos\pi(\bar{z}-\bar{z}_c)}{\cosh\pi(\bar{x}-\bar{x}_c)-\cos\pi(\bar{z}+\bar{z}_c)}\right],$$
$$-\infty<\bar{x}<\infty,\ \ 0\le\bar{z}\le 1,\ \ -\infty<\bar{x}_c<\infty,\ \ 0<\bar{z}_c<1. \tag{10}$$

The 2D limiting current density, $J_{max}$, is the value of $J$ beyond which there is no solution to the integral equation (9). It is clear from Eqs. (9) and (10) that the dimensionless quantity, $J_{max}/J_{CL}$ , depends only on the two normalized parameters, $\bar{w}$ and $\Delta$. Since $J_{max}/J_{CL}$ may become arbitrarily large as $\bar{w}\to 0$, we introduce another dimensionless parameter, $K_2=(2/9\pi)(w/d)(J/J_{CL})$, which remains finite in the limit $\bar{w}\to 0$. Note that $K_2$ is proportional to the total current transported to the anode per unit length in $y$ (Fig. 1; see also Eq. (3.11) of [39]). On an infinitesimally thin sheet, we set $\bar{x}=\bar{x}_c=0$, in which case the Green's function (10) is readily recognized to be Eq. (3.8) of [39]. In the limits $\bar{w}\to 0$ and $\bar{w}\to\infty$, we have verified that Eq. (9) reduces to Eqs. (3.10) and (3.4) of [39], respectively, where the integral equations are solved iteratively in these limits. Once more, $J_{max}/J_{CL}$ and the maximum value of $K_2$, denoted as $K_2(max)$, depend only on $\bar{w}$ and $\Delta$, since we have assumed $\bar{p}\to\infty$.

We may solve for $\bar{\phi}(\bar{x},\bar{z})$ iteratively from the integral equation (9) at fixed values of $J/J_{CL}$, $\bar{w}$ and $\Delta$. The iteration process begins with the vacuum potential, i.e., $\bar{\phi}^{(0)}(\bar{x},\bar{z})=\bar{z}$ for $k=0$. At the $k$-th iteration, $k=1,2,3,\dots$, the potential distribution $\bar{\phi}^{(k)}(\bar{x},\bar{z})$ may be obtained from the RHS of Eq. (9) in which $\bar{\phi}(\bar{x}_c,\bar{z}_c)$ is approximated by $\bar{\phi}^{(k-1)}(\bar{x}_c,\bar{z}_c)$. We consider $J_{max}/J_{CL}$ is reached if this iteration converges at a slightly lower value of $J/J_{CL}$ but does not converge at a slightly higher value of $J/J_{CL}$. In Appendices B and C, we outline the derivations of the scaling laws Eqs. (3) and (4), respectively, obtained from this iterative procedure. These two asymptotic formulas are compared with the differential equation approach (POISSON) and with PIC simulation results (XOOPIC) in Fig. 2 for the ten test cases, E1 – E10.

Appendix C points out that when $2\pi\Delta\ll\bar{w}$, in which case Eq. (3) applies, the term $\partial E_z/\partial z$ is dominant in the Poisson equation near the potential minimum as $J\to J_{max}$, and that when $\bar{w}\ll 2\pi\Delta$, in which case Eq. (4) applies, the term $\partial E_x/\partial x$ is dominant in the Poisson equation near the potential minimum as $J\to J_{max}$. Figure 3 shows that this is indeed the case in the very high-resolution POISSON runs of Tests E10 and E7.

Appendix B shows that Eq. (3) may also be derived using the idea introduced by Luginsland et al. [10] and Lau [35], viz, by equating the magnitude of the vacuum electric field $(V/d)$ with the magnitude of the self-electric field evaluated at the center line of the emission stripe that is due to the space charge within the AK gap. Here, this self-electric field includes all image charge effects but, for the present manuscript, we assume that the electron motion is subjected only to the vacuum electric field. The latter assumption of using the vacuum electric field as an approximation was motivated by our earlier results [39][40] that $K_2(max)$ becomes vanishingly small in the limit $(\Delta,\bar{w})\to(0,0)$. The use of this assumption also leads to the next order correction to Eq. (3), given in Eq. (B9) of Appendix B where it is argued that this assumption is valid to a good approximation.

The approach given in the preceding paragraph also yields the approximate value of $J_{max}/J_{CL}$ for a circular emission patch of radius $R$, $R\ll d$. The result is presented as Eq. (5) which is also derived in Appendix B, where the 2-term expansion of $J_{max}/J_{CL}$ is given by Eq. (B10). Using Eq. (5), the maximum current, $I_{max}=J_{max}\times\pi R^2\cong 9.72\times 10^{-6}A\times[V(volt)]^{3/2}(R/d)^{3/2}$, shows the $(ER)^{3/2}$ scaling (known for the "cigar regime" in the accelerator / photoinjector community [19][20][21]) where $E=V/d$ is the vacuum electric field. Indeed, this expression for $I_{max}$ may be shown to be identical to Eq. (6) of Denham et al. [20], in the example following that equation. Denham et al. reported their first experimental confirmation of this $(ER)^{3/2}$ scaling. Schamuilov et al. [21] examined the connection between this $(ER)^{3/2}$ scaling with Luginsland et al. [10] and Lau [35].

There are several intrinsic inconsistencies in the derivation of the scaling laws, Eqs. (3) and (5), in addition to the use of the vacuum electric field in the integrand. Toward the end of Appendix B, we argue that these inconsistencies appear to be not serious, especially since the resulting scaling laws compare well with POISSON code results and with experimental data.

**Particle-in-Cell Simulation**

The 2D–3V Particle-In-Cell (PIC) code XOOPIC [64] is used to perform electrostatic simulations on the single-period geometry shown in Fig. 1. It shows the temporal-spatial evolution in simulations, and it captures the kinetic effects contributing to virtual cathode formation not included in either the differential or the integral equation approaches. The perfect electric conductor (PEC) boundaries at $z = 0$ and at $z = d$, and the periodic boundary conditions at $x = \pm p/2$ are imposed. A uniform axial magnetic field of $100$ T along the $z$-direction is applied to suppress transverse electron motion while retaining a fully two-dimensional field solution [10]. We fix $V = 2.5$ kV, $d = 1$ cm, and vary $(\bar{w}, \Delta, \bar{p})$ in Tests E1–E10 as in POISSON. We compare the XOOPIC and POISSON results of these Tests in Fig. 2.

The domain is discretized using a uniform Cartesian grid with $n_z$ cells along $z$ and $n_x$ cells along $x$. The time integration employs an explicit standard PIC scheme with time step $\Delta t$ satisfying the Courant–Friedrichs–Lewy stability condition. In all cases, the spatial resolution is chosen to resolve the gap spacing $d$, the normalized emission width $\bar{w}$, and to accurately capture the position and depth of the electrostatic potential minima.

For Tests E1 ($\bar{w} = 0.1, \bar{p} = 4$) and E2 ($\bar{w} = 0.05, \bar{p} = 2$), corresponding essentially to very widely separated electron sheets since $\bar{p} \gtrsim 2$, we take $n_z = 1400$ and $n_x = 1600$, providing approximately 40 cells across the beam width. These runs are advanced to approximately $10$ ns and require approximately 3 hours of wall-clock time on a single compute node. This was sufficient to eliminate transient effects associated with particle injection and initial field adjustment. Tests E3–E6 ($\bar{w} = 0.01, \bar{p} = 2$) use $n_z = 1000$ and $n_x = 2000$ and are typically run to $100$ ns. For Tests E7–E10 ($\bar{w} = 0.001, \bar{p} = 2$), the transverse resolution is increased to $n_x = 20{,}000$; in these cases, computational constraints limit the longitudinal resolution to $n_z = 200$ cells across the gap, reducing the accuracy of the extracted potential minimum, particularly at low emission energies with a very small value of $\Delta$. The resulting increase in particle count leads to substantially longer runtimes, of order 72 hours for around $100$ ns of physical time. This low resolution ($n_z = 200$) is one reason for the discrepancy between XOOPIC and POISON for Tests E7 - E10 shown in Fig. 2. In general, a lower value of $n_z$ in XOOPIC leads to much greater uncertainties in the value of the potential minimum, as shown in Fig. 4b.

Grid convergence is verified by increasing $n_x$ and $n_z$ until negligible changes are observed in the electrostatic potential, particle trajectories, and current limit, while temporal convergence is confirmed by reducing $\Delta t$. Electron emission is modeled by injecting macroparticles with a weight corresponding to $2 \times 10^7$ physical electrons per unit length in the $y$-direction, which is the ignorable coordinate, from the emitting strip of width $w$, with prescribed current density $J$ and initial velocity $v_{in}$. As $J$ increases, the simulations transition from steady transport to a time-dependent regime marked by the formation of a local axial potential minimum near the beam centerline, partial electron reflection toward the cathode, and sustained oscillations in the electric field, current, and charge density. This transition consistently initiates around $x = 0$, while the transverse edges remain less strongly field-reversed at onset. Figure 4 shows the spatial-temporal evolution for the potential and density profiles for Test E3 when the injection current density $J$ is slightly below and slightly above $J_{crit}$. Some observed features of virtual cathode formation are shared with previous 1D models [46]-[50], and 2D simulations [10][23][25].

# IV. DISCUSSION

The major results have been summarized toward the end of Section II and will not be repeated here. This section examines various issues. To within the uncertainties of the theoretical models, we now have the scaling laws for 2D CLL for an isolated emitting patch of an *arbitrary* size, assuming $\Delta = 0$. The 2D CLL for an isolated electron stripe is approximately given by Eq. (2a) for $w/d > 0.1$, and by Eq. (3) for $w/d < 0.1$. At $w/d = 0.1$, where both scaling laws are expected to be the least accurate, their values differ only by 8.1 percent. Likewise, the 2D CLL for an isolated circular emitting patch is approximately given by Eq. (2b) for $R/d > 0.5$, and by Eq. (5) for $R/d < 0.5$. At $R/d = 0.5$, their values differ by 22 percent.

It is a bit surprising that when the thin electron sheets are separated by more than $1/2$ of the AK gap separation, the maximum anode current per electron sheet may be reliably calculated as if the electron sheets were widely separated, as demonstrated by the solid curves in Fig. 5 when $\bar{p} = p/d > 1/2$. We note here that the number $1/2$ may, at times, be regarded as a sufficiently "large" expansion parameter for asymptotic analysis. For example, the Bohr-Sommerfeld quantization rule is known to give a surprisingly accurate prediction of the eigenvalue for the ground state, which has only $1/2$ wavelength (instead of many wavelengths) between the "turning points" of the eigenfunction. One explanation is that $1/2$ wavelength, after being multiplied by the numerically large factor $2\pi$, corresponds to a phase of $\pi$ radian, and 3.1416 is a sufficiently large expansion parameter for asymptotic analysis of phase integral method, from which the quantization rule is derived. Another example of similar accuracy in the phase integral method may be found in [65]. We note that Stirling's asymptotic formula for the factorial function $\Gamma(n+1) = n!$ is also remarkably accurate for $n \geq 1/2$. [66]

The present study assumes uniform emission current density over a small emission patch and a low electron emission velocity. We believe that the scaling law for the total maximum current drawn from this small patch is applicable even when this assumption is replaced by the assumption of space-charge-limited emission over the entire emission patch. This issue was examined in depth for the case of an infinitesimally thin sheet [39], for both zero and nonzero emission velocity, as in this $w = 0$ limit, there can be no distinction whether the electron emission is uniformly emitted over an infinitesimally thin sheet, or emitted under the space-charge-limited condition over the same sheet even for the case $\Delta \to 0$. We simply find the condition on the sheet current beyond which there is no solution to the integral equation.

We have also assumed an infinite magnetic field in the $z$-direction. There is strong evidence that the total current transported to the anode is insensitive to the presence or absence of such an imposed magnetic field, either for sparsely separated patches of electron emission [10][11][37], or for a cathode with randomly distributed emission patterns on small emission patches [37][38][44]. This is consistent with the notion that the "1D motion" assumption, while helpful to theory, is not very restrictive for nonmagnetized diodes in practice.

We assume that all electrons are emitted with the same velocity. If a thermal distribution in the emission velocity is assumed, the analysis will be far more complex, and the scaling law may change quantitatively, though not qualitatively. This somewhat optimistic conjecture is also based on our prior studies [37]-[39] which show that the assumption of constant and uniform emission velocity yield results on the extractable anode current qualitatively similar (a) to the experimental results on thermionic cathodes, (b) to authentic PIC simulations, and

(c) to the semi-analytic model; in which the experimentally determined work function distribution is incorporated into the Richardson-Dushman thermionic emission law for each of the randomly distributed (small) emission patches.

A related question is whether we need to accurately resolve the potential minimum in order to reliably predict the extractable anode current. We are unable to definitively answer this question at this point. Our tentative answer is a guarded "no", for the important regime of small but finite $\overline{w}$ and $\Delta$ that is the focus of this paper. One reason for this speculation is that in this case, the virtual cathode is very close to the cathode surface, and the value of the potential at the potential minimum is very small compared with the AK gap voltage. Thus, the 1D Child-Langmuir law, which has been used as the yardstick throughout this study, would not be affected by much. By extension, neither would the maximum anode current from randomly distributed emission patches on the cathode surface that was examined in detail in [37][38]. But for an individual, isolated emission patch, the scaling laws for the regimes $2\pi\Delta \ll \overline{w}$ and $\overline{w} \ll 2\pi\Delta$, might have a different form from Eqs. (3) and (4) and would depend on the emission model, where $\Delta$ is now a measure of the initial emission velocity also dependent on that emission model. Another indication for the "robustness" in the anode current, despite some intrinsic numerical uncertainties, may be inferred from Figs. 2b,c,d. These figures show excellent agreement between the scaling laws and the very high resolution runs of POISSON in several test cases which span over a wide range of $\overline{w}$ and $\Delta$. As noted toward the end of Appendix B, there are several intrinsic inconsistencies in the derivation of the scaling law (3). Yet Eq. (3) still provides excellent agreement with the very high resolution POISSON code results, as demonstrated in Fig. 2 despite these inconsistencies. Even more significantly, the extension of Eq. (3) to a circular emitting region, Eq. (5), is recently confirmed in photocathode experiments [20], despite similar built-in inconsistencies in the derivation of Eq. (5). Our POISSON code results suggest that, for the difficult regime of small $(\overline{w}, \Delta)$, $J_{max}$ can be reliably computed if there are more than 3 numerical cells from the location of the potential minimum to the cathode (in z, as shown in the row displaying $z_{min}/d_z$ in Table 1), and to the beam edge (in x, as shown in the example of Fig. 3d).

The present study may be extended in many directions, including the effects of cathode surface roughness, different emission mechanisms, nonplanar diodes, the quantum regimes, time-dependent emission profiles, initiation of electrical breakdown, the presence of multiple species, and the effects of a transverse magnetic field. [6]

## ACKNOWLEDGEMENTS

This work was supported in part by the Air Force Office of Scientific Research (AFOSR) under Grant numbers FA9550-20-1-0409, FA9550-21-1-0184, FA9550-22-1-0523, and FA9550-26-1-B352, and in part by the National Science Foundation (NSF) under Grant number 2516752.

## DATA AVAILABILITY

The data that support the findings of this study are available within the article [and its supplementary material].

# APPENDIX A. DERIVATION OF INTEGRAL EQUATION EQ. (9)

Assuming an infinite magnetic field in the z-direction so that all electron motions are restricted to the z-direction, with the same initial energy, $E_{in}$ , the electron velocity $v(x,z)$ is given by $mv^2/2 = E_{in} + e\phi(x,z)$, yielding the charge density with a magnitude

$$\rho(x,z) = \frac{J}{v(x,z)} = J/[(2/m)(E_{in} + e\phi(x,z)]^{1/2} \tag{A1}$$

where $J$ ($> 0$) is the current density along the $z$-axis. For uniform emission current density over a single, isolated emission stripe of width $w$, $J$ is a constant, and the potential function $\phi(x,z)$ within the gap may be expressed as a superposition of the vacuum potential, $Vz/d$, and the space charge potential. This leads to the integral equation, (see Fig.1)

$$\phi(x,z) = Vz/d + \int_0^d dz_c \int_{-w/2}^{w/2} dx_c \frac{J}{[(2/m)(E_{in} + e\phi(x_c,z_c)]^{1/2}} G(x,z;x_c,z_c). \tag{A2}$$

In Eq. (A2), the Green's function $G(x,z;x_c,z_c)$ is the electrostatic potential at position $(x,z)$ due to a line charge of unit line charge density at position $(x_c,z_c)$, all within the two grounded plates (i.e., at zero potential) at $z = 0$ and $z = d$.

The Green's function $G(x,z;x_c,z_c)$ may be obtained from the image method. For the line charge located at $(x_c,z_c)$ within the grounded plates at $z = 0$ and $z = d$, there is an infinite series of image line charges located at $(x_c, 2nd + z_c)$, $n = \pm 1, \pm 2, \ldots$ . Including the contributions from these image charges, the electrostatic potential at position $(x,z)$ is the Green's function,

$$G(x,z;x_c,z_c) = \frac{1}{2\pi\epsilon_0}\sum_{n=-\infty}^{\infty} ln\left(\frac{L_{1n}}{L_{2n}}\right) = \frac{1}{4\pi\epsilon_0} ln\left[\frac{cosh\,\pi(\bar{x}-\bar{x}_c) - cos\,\pi(\bar{z}-\bar{z}_c)}{cosh\,\pi(\bar{x}-\bar{x}_c) - cos\,\pi(\bar{z}+\bar{z}_c)}\right],$$
$$-\infty < x < \infty, \qquad 0 \le z \le d, \qquad -\infty < x_c < \infty, \qquad 0 < z_c < d, \tag{A3}$$

$$L_{1n} = \sqrt{(x-x_c)^2 + [(2nd - (z-z_c)]^2}, \tag{A4a}$$

$$L_{2n} = \sqrt{(x-x_c)^2 + [(2nd + (z+z_c)]^2}\,. \tag{A4b}$$

Upon substituting Eq. (A3) into Eq. (A2), and using the dimensionless variables $\bar{\phi} = \phi/V$, $\bar{z} = z/d$, $\bar{z}_c = z_c/d$, $\bar{x} = x/d$, $\bar{x}_c = x_c/d$, $\bar{w} = w/d$, and $\Delta = E_{in}/\mathrm{e}V$, we obtain Eq. (9) of the main text. We remark that by setting $x = x_c = 0$ in Eq. (A3), the Green's function expressions become identical to Eq. (3.8) of [39] where an analogous derivation is given.

# APPENDIX B. DERIVATION OF LIMITING CURRENT FOR $\Delta = 0$ AND SMALL $w/d$

We set $\Delta = 0$. In the limit of infinitesimally thin sheet ($\bar{w} = w/d \to 0$), the current density profile becomes $K_2\delta(x)$ where $\delta$ represents the Dirac delta function. The maximum value of $K_2$, $K_2(max)$, equals zero; a result proven in [39]. Thus, when $\bar{w}$ is very small but finite, we anticipate that $K_2(max)$ will also be very small but finite, in which case $K_2(max)$ may be obtained after just one iteration ($k = 1$) in the iterative solution of the integral equation (9). At $k = 1$, the potential distribution, $\bar{\phi}^{(1)}(\bar{x},\bar{z})$, is obtained by using the vacuum potential, $\bar{\phi}^{(0)}(\bar{x},\bar{z}) = \bar{z}$ in the RHS of Eq. (9),

$$\bar{\phi}^{(1)}(\bar{x},\bar{z}) = \bar{z} + \frac{2}{9\pi}\left(\frac{J}{J_{CL}}\right)\int_0^1 d\bar{z}_c \int_{-\bar{w}/2}^{\bar{w}/2} d\bar{x}_c \frac{1}{\bar{z}_c{}^{1/2}} \bar{G}(\bar{x},\bar{z};\bar{x}_c,\bar{z}_c)\,, \quad 0 \le \bar{z} \le 1\,, -\infty < \bar{x} < \infty. \tag{B1}$$

The maximum value, $J_{max}/J_{CL}$, is reached when the total electric field vanishes on the center line of the emission stripe at the cathode surface [35] for $\Delta = 0$. Using Eq. (B1), this condition is approximately given by,

$$\frac{\partial\bar{\phi}^{(1)}(\bar{x},\bar{z})}{\partial\bar{z}} = 0, \qquad \text{at } \bar{x} = 0, \bar{z} = 0, \tag{B2}$$

yielding,

$$0 = 1 - \frac{4}{9}\left(\frac{J_{max}}{J_{CL}}\right)\int_0^1 d\bar{z}_c \int_0^{\bar{w}/2} d\bar{x}_c \frac{1}{\bar{z}_c{}^{1/2}} \frac{\sin\pi\bar{z}_c}{\cosh\pi\bar{x}_c - \cos\pi\bar{z}_c}\,. \tag{B3}$$

For $\bar{w} \ll 1$, we may approximate the factor $\cosh\pi\bar{x}_c \cong 1 + (\pi\bar{x}_c)^2/2$ in Eq. (B3), in which case the second integral (in $\bar{x}_c$) may be explicitly carried out, and Eq. (B3) may be rewritten

$$0 = 1 - \frac{4}{9}\left(\frac{J_{max}}{J_{CL}}\right)\left(\frac{2}{\pi}\right)^{3/2} \int_0^{\pi/2} dx \frac{\cos x}{x^{1/2}} \tan^{-1}\left(\frac{\pi\bar{w}/4}{\sin x}\right)\,. \tag{B4}$$

Taking the leading term of the asymptotic expansion for the integral in Eq. (B4) for $\bar{w} \ll 1$, we obtain,

$$\frac{J_{max}}{J_{CL}} \cong \frac{9}{8} \times \frac{1}{\sqrt{\bar{w}}}, \quad \bar{w} \ll 1, \qquad \Delta = 0, \tag{B5}$$

$$K_2(max) \equiv \left(\frac{2}{9\pi}\right)\left(\frac{w}{d}\right)\left(\frac{J_{max}}{J_{CL}}\right) \cong \frac{1}{4\pi}\sqrt{\bar{w}}, \quad \bar{w} \ll 1, \qquad \Delta = 0. \tag{B6}$$

Equation (B5) is the scaling law Eq. (3) of the main text, and the corresponding maximum value of $K_2$ is given by Eq. (B6). Note that $K_2(max) \to 0$ as $\bar{w} \to 0$, as expected [39].

An alternate derivation of Eq. (B5) without using the integral equation follows, as suggested by Luginsland et al. [10] and Lau [35]. Since $K_2(max)$ is small for small $\bar{w}$, we may assume that the electrons in this thin electron sheet are primarily subjected to the vacuum electric field. Now, if all electrons are subjected only to the vacuum electric field, $V/d$, their charge density is [cf. Eq. (A1)]

$$\rho(x,z) = \frac{J}{v(x,z)} = J/[(2/m)(eVz/d)]^{1/2}. \tag{B7}$$

Using this approximation, the 2D limiting current density, $J_{max}$, may be calculated by equating the magnitude of the vacuum electric field ($V/d$) with the magnitude of the self-electric field at the center line on the cathode surface due to the space charge density $\rho(x,z)$ that is given by Eq. (B7). This yields,

$$\frac{V}{d} = \frac{2J_{max}d}{\pi\epsilon_0\sqrt{2eV/m}}\int_0^1 d\bar{z}_c \int_0^{\bar{w}/2} d\bar{x}_c \frac{1}{\bar{z}_c^{\,1/2}}\left\{\frac{\bar{z}_c}{\bar{x}_c^{\,2}+\bar{z}_c^{\,2}} + \sum_{n=1}^{\infty}\left[\frac{2n+\bar{z}_c}{\bar{x}_c^{\,2}+(2n+\bar{z}_c)^2} - \frac{2n-\bar{z}_c}{\bar{x}_c^{\,2}+(2n-\bar{z}_c)^2}\right]\right\}. \quad \text{(B8)}$$

The space charge electric field at $(x,z) = (0,0)$ is given by the RHS of Eq. (B8). In the curly bracket of Eq. (B8), the first term accounts for the self-electric field due to the space charge [Eq. (B7)] between the grounded plates at $\bar{z} = 0$ and $\bar{z} = 1$; and the remaining terms in the infinite sum account for electric field due to the infinite series of image charges. We find that the first term in the curly bracket of Eq. (B8) gives the dominant contribution when $\bar{w} \ll 1$, and that this dominant contribution arises from the immediate vicinity of $\bar{z}_c = 0$ in the $\bar{z}_c$-integral of Eq. (B8). That is, most of the self electric field is due to the space charge in the immediate vicinity of the cathode (i.e., $\bar{z}_c \ll \bar{w}$ ) when $\bar{w} \ll 1$, $\Delta = 0$. Thus, in this regime ($\bar{w} \ll 1$, $\Delta = 0$), $|\partial E_z/\partial z| \gg |\partial E_x/\partial x|$ near the virtual cathode as $J \to J_{max}$.

We next perform a 2-term asymptotic expansion for the double integral in Eq. (B8) for $\bar{w} \ll 1$. This leads to the 2-term expansion for $J_{max}/J_{CL}$,

$$\frac{J_{max}}{J_{CL}} \cong \frac{9}{8} \times \frac{1}{\sqrt{\bar{w}}}\left(1 + 0.413\sqrt{\bar{w}}\right), \quad \bar{w} \ll 1, \qquad \Delta = 0, \quad \text{(B9)}$$

where the leading term is the same as Eq. (B5), or Eq. (3) of the main text. As a check, we record that the fractional difference in $J_{max}/J_{CL}$ between the asymptotic formula Eq. (B9) and the numerical solution to Eq. (B4) is (0.017, 0.0085, 0.0017, 0.00016, 0.000014) for $\bar{w}$ = (0.1, 0.05, 0.01, 0.001, 0.0001), respectively.

We have used a similar technique to obtain the 2-term expansion of $J_{max}/J_{CL}$ for an isolated, circular emitting region of radius $R \ll d$, assuming uniform emission current density over the emission region and zero initial electron velocity. We obtain, analogous to Eq. (B9),

$$\frac{J_{max}}{J_{CL}} \cong \frac{1.328}{\sqrt{R/d}}\left(1 + 0.34\left(\frac{R}{d}\right)^{3/2}\right), \quad R/d \ll 1, \qquad \Delta = 0. \quad \text{(B10)}$$

While Eq. (B10) is derived assuming $\Delta = 0$, its independence of $\Delta$ suggests that the scaling in Eq. (B10) is also valid for a small but finite $\Delta$, provided $\Delta \ll R/d \ll 1$. The maximum current, $I_{max}$, then reads,

$$I_{max} \equiv J_{max} \times \pi R^2 \cong 9.72 \times 10^{-6} A \times [V(volt)]^{3/2}(R/d)^{3/2}\left(1 + 0.34(R/d)^{3/2}\right), \quad \Delta \ll R/d \ll 1, \quad \text{(B11)}$$

where we have used the numerical value in $J_{CL}$. The leading term in Eq. (B10) is displayed as Eq. (5) of the main text. Note that the leading term of Eq. (B11) shows the well-known scaling of $I_{max} \propto (ER)^{3/2}$ in the "cigar regime" in the photoinjector community [19][20][21], where $E = V/d$ is the vacuum electric field, and this scaling is recently experimentally verified, for the first time, for photoinjectors [20].

It is worth pointing out that there are intrinsic inconsistencies in the derivation of the scaling laws (B9), (B10) and (B11). When calculating the self-electric field, we have assumed that the space charge effect is unimportant so that the electron trajectory is subjected only to the vacuum electric field. Then we derive Eqs. (B9), (B10), and (B11) by demanding that the self-electric field is very important after all, so important that it exactly cancels the vacuum electric field at the potential minimum. Since the self-field is primarily due to the space charge within a very narrow region from the cathode ($\bar{z}_c \ll \bar{w} \ll 1$) for $\Delta = 0$, this inconsistency may not be as serious. This is perhaps why the leading term of the scaling law (B11) is verified by the recent experiments on photoinjectors by Denham et al. [20]. However, this inconsistency in the scaling laws, (B9), (B10) and (B11), is always present. Another source of inconsistency is that, in our iterative solutions in both the differential and integral equation approaches, we used the non-convergence of the iterative solutions as the condition to determine $J_{max}/J_{CL}$, and this non-convergence condition is also used in Lau et al. [39] and Chernin et al. [40]. *That is, the iterative* *solutions to the integral equation, and to the differential equation that yields the POISSON code results, did not use the condition of zero electric field on the cathode surface as the condition to determine $J_{max}/J_{CL}$.* The third source of inconsistency arises when Eq. (B9), or Eq. (3), is used to compare with the POISSON code results. Equations (B9) and (3) have assumed that the potential minimum is located on the cathode surface, $z = 0$. For nonzero $\Delta$, which is always assumed in the POISSON code, the potential minimum is necessarily located at $z = z_{min} > 0$, as shown in the listed values of $z_{min}$ in Table 1 of Appendix D. So, some discrepancies between the asymptotic theory and the numerical result from POISSON are unavoidable. Yet Figs. 2b,c,d show remarkable agreement between POISSON and Eq. (3) despite these intrinsic inconsistencies in our analytic derivation of the scaling law, Eq. (3).

Since the scaling law (B9) is independent of $\Delta$ and it was derived by assuming $(\Delta, \bar{w})$ = (0, small but finite), we will postpone to Appendix C to argue that Eq. (B9) is expected to be applicable for nonzero $\Delta$ provided $\Delta \ll \bar{w}/2\pi$ [cf. Eq. (3)]. In Appendix C, we consider the other regime, assuming $(\Delta, \bar{w})$ = (small but finite, 0), whose scaling law, Eq. (4), is expected to be applicable if $\Delta \gg \bar{w}/2\pi$.

## APPENDIX C. DERIVATION OF LIMITING CURRENT FOR $w/d = 0$ AND SMALL $\Delta$

We set $\bar{w} = w/d = 0$. The emission stripe is infinitesimally thin. This special case was treated in detail in [39]. From a careful iterative solution to the reduced integral equation, the following empirical formula was synthesized (cf. Fig. 6 of [39])

$$K_2(max) \cong 0.2336 \times \Delta^{0.5274}, \quad 10^{-5} \leq \Delta \leq 10^{-2}, \qquad \bar{w} = 0. \tag{C1}$$

Upon using the definition, $K_2 \equiv (2/9\pi)(w/d)(J/J_{CL})$, the scaling law Eq. (4) for $J_{max}/J_{CL}$ follows.

Since Eq. (C1) is derived in the regime $(\Delta, \bar{w})$ = (small but finite, 0), whereas Eq. (B6) is derived in the regime $(\Delta, \bar{w})$ = (0, small but finite), these two asymptotic formulas assume the same numerical value when

$$\bar{w} = 8.6172 \times \Delta^{1.0548}. \tag{C2}$$

For convenience, over the range of $\bar{w} \in [0.001, 0.1]$, which is the main focus of this paper, instead of using the cumbersome Eq. (C2), we will replace it with a much simpler, and more appealing approximation,

$$\bar{w} = 2\pi\Delta. \tag{C3}$$

Thus, the scaling law (C1) applies if $\bar{w} \ll 2\pi\Delta$, and the scaling law (B6) applies if $\bar{w} \gg 2\pi\Delta$. This is the reason for the inequalities that appear in Eqs. (3) and (4) of the main text. In Fig. 2, the red vertical lines represent Eq. (C3). They provide an excellent approximation to Eq. (C2), which gives the value of $\bar{w}$ at the intersection of the two straight lines depicting Eqs. (3) and (4) in Fig. 2. Note further that the factor $2\pi$ in Eq. (C3) is a large numerical factor; it places Test E3 with $\bar{w} = \Delta = 0.01$ in the regime, $\bar{w} \ll 2\pi\Delta$, so that the scaling law Eq. (4) applies. See Fig. 2(a).

The following features may be noted [39], on the iterative solutions of the integral equation that led to the scaling law, Eq. (4). (a) Multiple iterations ($k > 1$) are required to compute $K_2(max)$. This means that the electron trajectory is not subjected only to the vacuum field. (b) The potential minimum at $K_2(max)$ is located at some distance from the cathode surface; we have no formula for the location of the potential minimum, nor for its value, as a function of $\Delta$. Their normalized numerical values, $z_{min}/d$ and $\phi_{min}/V$, obtained from the POISSON code, are listed in Table 1 in Appendix D. (c) Equation (C1) is derived for an infinitesimally thin sheet, which we have argued that it is also applicable for a thin sheet of finite width provided $\bar{w} \ll 2\pi\Delta$. For such a very thin sheet, we then anticipate $|\partial E_z/\partial z| \ll |\partial E_x/\partial x|$ near the virtual cathode as $J \to J_{max}$. This is opposite to the case studied in Appendix B. (d) In other words, when $\bar{w} \ll 2\pi\Delta$, the term $\partial E_x/\partial x$ is dominant in the Poisson equation near the potential minimum as $J \to J_{max}$. When $\bar{w} \gg 2\pi\Delta$, the term $\partial E_z/\partial z$ is dominant in the Poisson equation near the potential minimum as $J \to J_{max}$. The very high-resolution POISSON data show that this is indeed the case, as exemplified by the two Tests, E7 and E10, results from which are illustrated in Fig. 3.

# APPENDIX D. POISSON CODE DATA FOR TESTS E1-E10, AND FOR ADDITIONAL TESTS IN FIGURE 5

Table 1 lists the POISSON code parameters that we used for the finest resolution runs for Tests E1 – E10. It also lists the parameters used in the eight (8) POISSON data points that are marked by the crosses in Fig. 5. The values of $< \bar{J}_{max} >$ for these eight data points are listed in the eight entries in the last row of this Table; see Note 5 at the end of this appendix.

**TABLE 1.** POISSON simulation parameters and calculated limiting-current quantities for Tests E1-E10 (top part) and for additional tests with smaller values of *p/d* (bottom part).

| V | $d$ | $J_{CL}$ |
|---|---|---|
| 2.5 kV | 1.0 cm | 0.292 A/cm$^{-2}$ |

| Parameter | E1 | E2 | E3 | E4 | E5 | E6 | E7 | E8 | E9 | E10 |
|---|---|---|---|---|---|---|---|---|---|---|
| $w/d$ | 0.1 | 0.05 | 0.01 | 0.01 | 0.01 | 0.01 | 0.001 | 0.001 | 0.001 | 0.001 |
| $p/d$ | 5.120 | 2.560 | 2.048 | 2.048 | 2.048 | 2.048 | 3.277 | 3.277 | 3.277 | 2.048 |
| $\Delta$ | 1.0E-02 | 1.0E-03 | 1.0E-02 | 1.0E-03 | 1.0E-04 | 1.0E-05 | 1.0E-02 | 1.0E-03 | 1.0E-04 | 1.0E-05 |
| $f_J$ | 1.349 | 1.099 | 1.349 | 1.099 | 1.030 | 1.010 | 1.349 | 1.099 | 1.030 | 1.010 |
| $E_{inj}/(e * \text{Volt})$ | 25.0 | 2.5 | 25.0 | 2.5 | 0.250 | 0.025 | 25.0 | 2.5 | 0.250 | 0.025 |
| $n_x$ | 2048 | 2048 | 4096 | 4096 | 4096 | 4096 | 32768 | 32768 | 32768 | 16384 |
| $n_z$ | 2000 | 1400 | 2000 | 2000 | 10000 | 50000 | 2000 | 8000 | 12000 | 100000 |
| $z_{min}/d$ | 1.900E-02 | 3.571E-03 | 1.200E-02 | 2.000E-03 | 4.000E-04 | 6.000E-05 | 9.500E-03 | 1.250E-03 | 2.500E-04 | 4.000E-05 |
| $z_{min}/d_z$ | 38 | 5 | 24 | 4 | 4 | 3 | 19 | 10 | 3 | 4 |
| $\phi_{min}/\text{V}$ | -4.701E-03 | -4.256E-04 | -5.784E-03 | -5.201E-04 | -3.526E-05 | -2.053E-06 | -5.686E-03 | -5.698E-04 | -4.888E-05 | -3.824E-06 |
| $J_{max}/(J_{CL} f_J)$ | 3.848 | 4.392 | 24.358 | 1.281 | 8.953 | 7.750 | 219.260 | 86.768 | 41.328 | 27.945 |
| $J_{max}/J_{CL}$ | 5.191 | 4.829 | 32.863 | 1.408 | 9.226 | 7.824 | 295.823 | 95.401 | 42.587 | 28.211 |
| $< \bar{J}_{max} >$ | | | 0.119 | | | | 0.0699 | 0.0265 | | |

| Parameter | $p/d \sim 0.1$ $\Delta = 10^{-2}$ | $p/d \sim 0.5$ $\Delta = 10^{-2}$ | $p/d \sim 0.1$ $\Delta = 10^{-3}$ | $p/d \sim 0.5$ $\Delta = 10^{-3}$ | $p/d \sim 2.0$ $\Delta = 10^{-3}$ |
|---|---|---|---|---|---|
| $w/d$ | 0.001 | 0.001 | 0.001 | 0.001 | 0.001 |
| $p/d$ | 0.102 | 0.512 | 0.102 | 0.512 | 2.048 |
| $\Delta$ | 1.0E-02 | 1.0E-02 | 1.0E-03 | 1.0E-03 | 1.0E-03 |
| $f_J$ | 1.349 | 1.349 | 1.099 | 1.099 | 1.099 |
| $E_{inj}/(e * \text{Volt})$ | 25.0 | 25.0 | 2.5 | 2.5 | 2.5 |
| $n_x$ | 2048 | 4096 | 1024 | 4096 | 16384 |
| $n_z$ | 2000 | 2000 | 8000 | 4000 | 4000 |
| $z_{min}/d$ | 1.800E-02 | 1.000E-02 | 1.625E-03 | 1.250E-03 | 1.250E-03 |
| $z_{min}/d_z$ | 36 | 20 | 13 | 5 | 5 |
| $\phi_{min}/\text{V}$ | -4.273E-03 | -5.501E-03 | -5.534E-04 | -5.557E-04 | -5.527E-04 |
| $J_{max}/(J_{CL} f_J)$ | 95.586 | 204.000 | 63.125 | 84.570 | 86.680 |
| $J_{max}/J_{CL}$ | 128.963 | 275.234 | 69.406 | 92.984 | 95.304 |
| $< \bar{J}_{max} >$ | 0.9372 | 0.3985 | 0.616 | 0.165 | 0.042 |

**Notes.**

1. $n_x$ is the number of cells over $-p/2 \leq x \leq p/2$; the POISSON calculation uses symmetry and solves $0 \leq x \leq p/2$.
2. $n_z$ is the number of cells across the diode gap, with $d_z = d/n_z$ being the cell size in z.
3. $z_{min}$ and $\phi_{min}$ are evaluated on the centerline $x = 0$.
4. $J_{max}/(J_{CL} f_J)$ is the maximum current density normalized by the one-dimensional Child-Langmuir value corrected by the Jaffe factor.
5. $< \bar{J}_{max} > = (w/p) J_{max}/(J_{CL} f_J)$ is the period-averaged normalized maximum current density. Only its value for the eight (8) cases marked by crosses in Fig. 5 is listed for clarity.

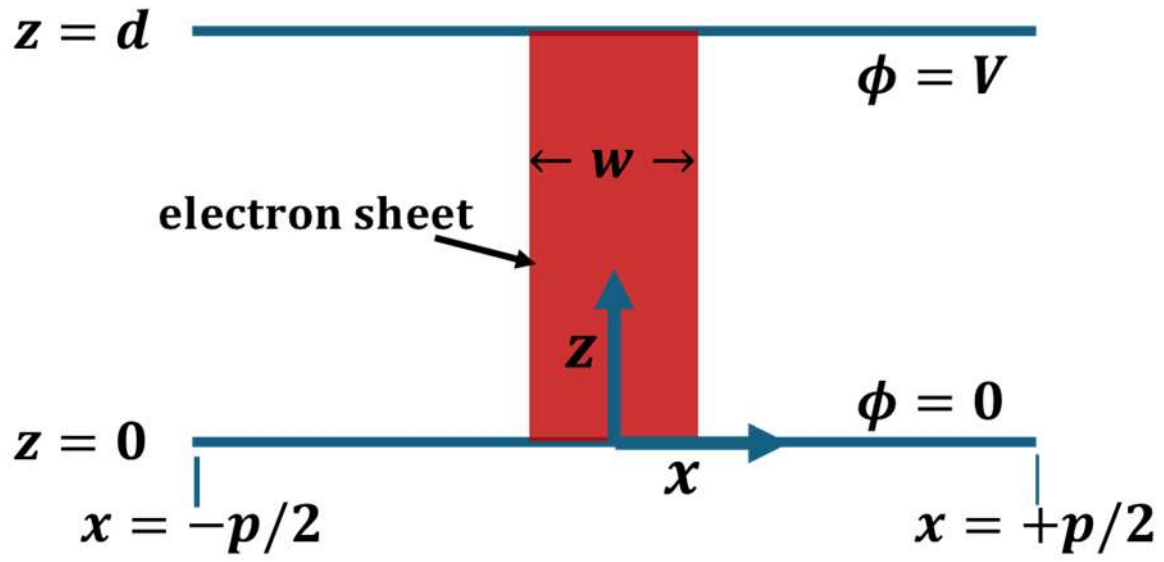

Fig. 1. A single period of an infinite array of period $p$ of electron sheets of width $w$ $(\leq p)$ in a planar diode of gap spacing $d$ and gap voltage $V$. Electrons are emitted from the surface at $z = 0$ with initial energy $E_{in}$, uniform current density $J$, and electron motion restricted to the $+z$-direction. The configuration is uniform and infinite in the $y$-direction, perpendicular to the page. The maximum value of $J$ is determined in terms of $V$, $d$, $w$, $p$, and $E_{in}$, using three (3) different approaches: differential equation, integral equation, and particle-in-cell simulation. The numerical results are compared, along results from new scaling laws that are constructed for the regime, $w/d \ll 1$, $E_{in}/\mathrm{e}V \ll 1$, and $p/d \gg 1$. Note that this model yields the maximum anode current of a periodic array of 2D emitters in terms of the packing density parameter, $p/d$, and that the results of this study are independent of the electron emission mechanism and cathode material.

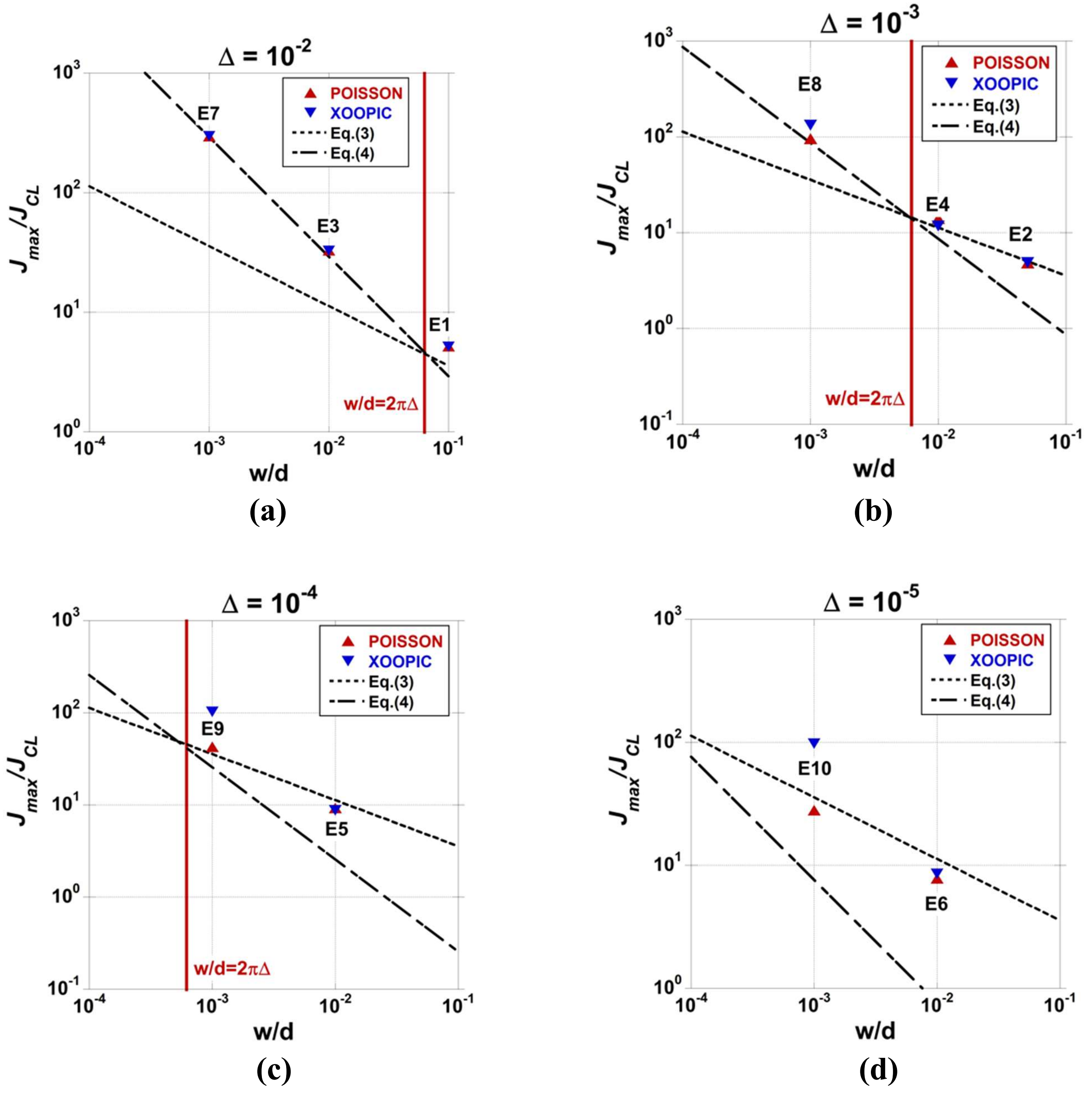

Fig. 2. Comparison of the 2D limiting current density from three (3) different approaches: differential equation (POISSON), particle-in-cell simulation (XOOPIC), and the scaling law Eqs. (3) and (4) obtained from the integral equation. Shown are the data for the normalized maximum current density, $J_{max}/J_{CL}$, as a function of $w/d \equiv \overline{w}$ at various $\Delta \equiv E_{in}/\mathrm{e}V$, for Tests E1-E10, grouped into (a) $\Delta = 10^{-2}$, (b) $\Delta = 10^{-3}$, (c) $\Delta = 10^{-4}$, and (d) $\Delta = 10^{-5}$. The red vertical line $\overline{w} = 2\pi\Delta$ separates the two regimes of the scaling laws. To the far left of this line ($\overline{w} \ll 2\pi\Delta$), Eq. (4) applies, and $\partial E_x/\partial x \gg \partial E_z/\partial z$ near the potential minimum. To the far right of this line ($\overline{w} \gg 2\pi\Delta$), Eq. (3) applies and $\partial E_x/\partial x \ll \partial E_z/\partial z$ near the potential minimum. All POISSON data for Tests E1-E10 have the potential minimum resolved. The XOOPIC data have progressively poorer resolutions as $\Delta$ decreases. The parameters used and the numerical results obtained from POISSON, for Tests E1 to E10, are listed in Table 1 in Appendix D.

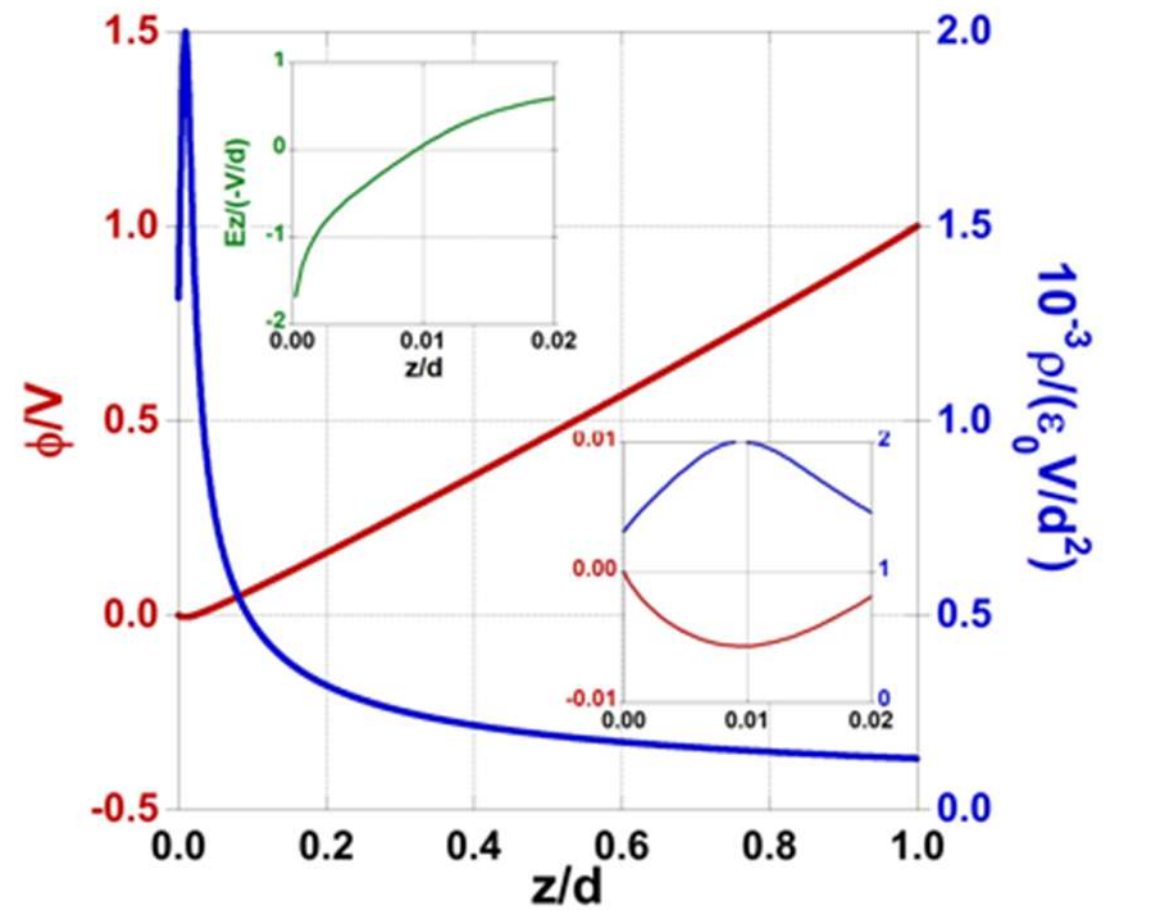

E7
(a)

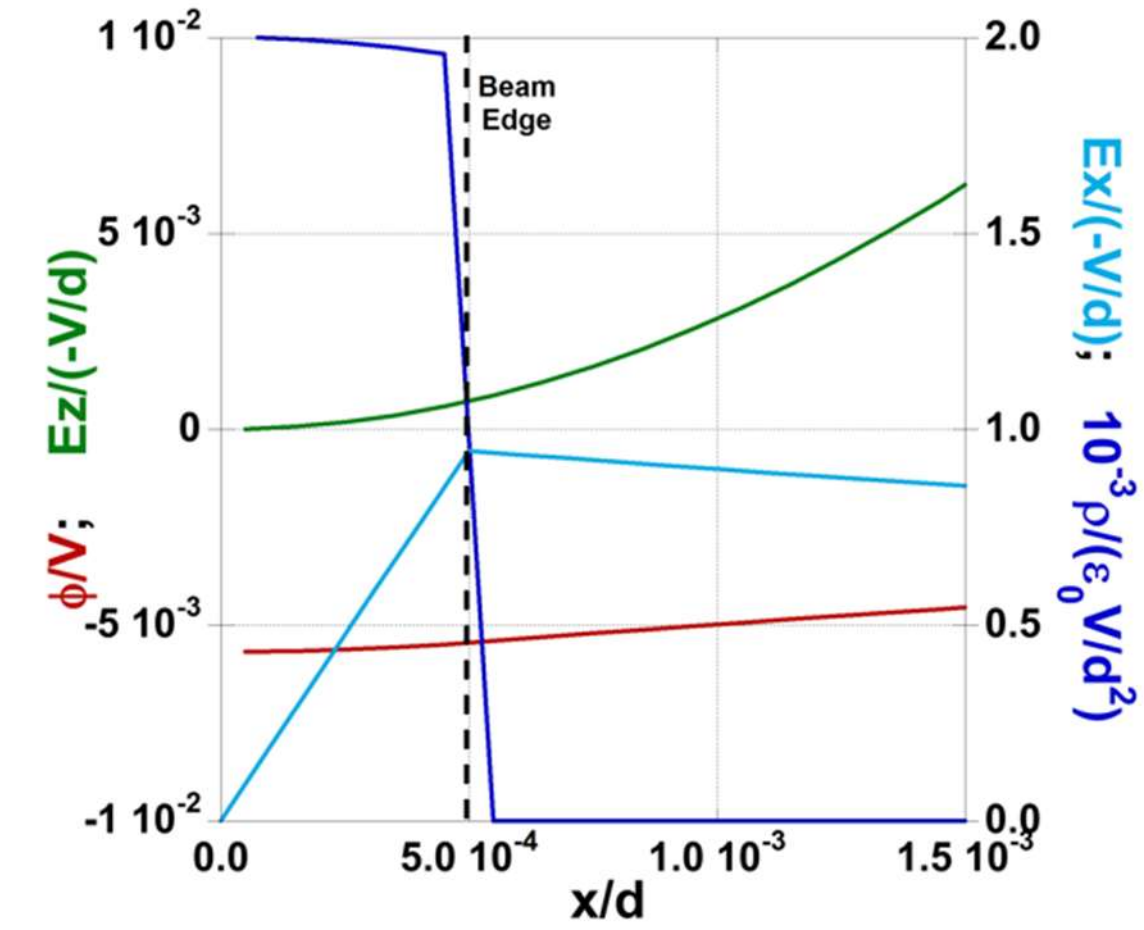

E7
(b)

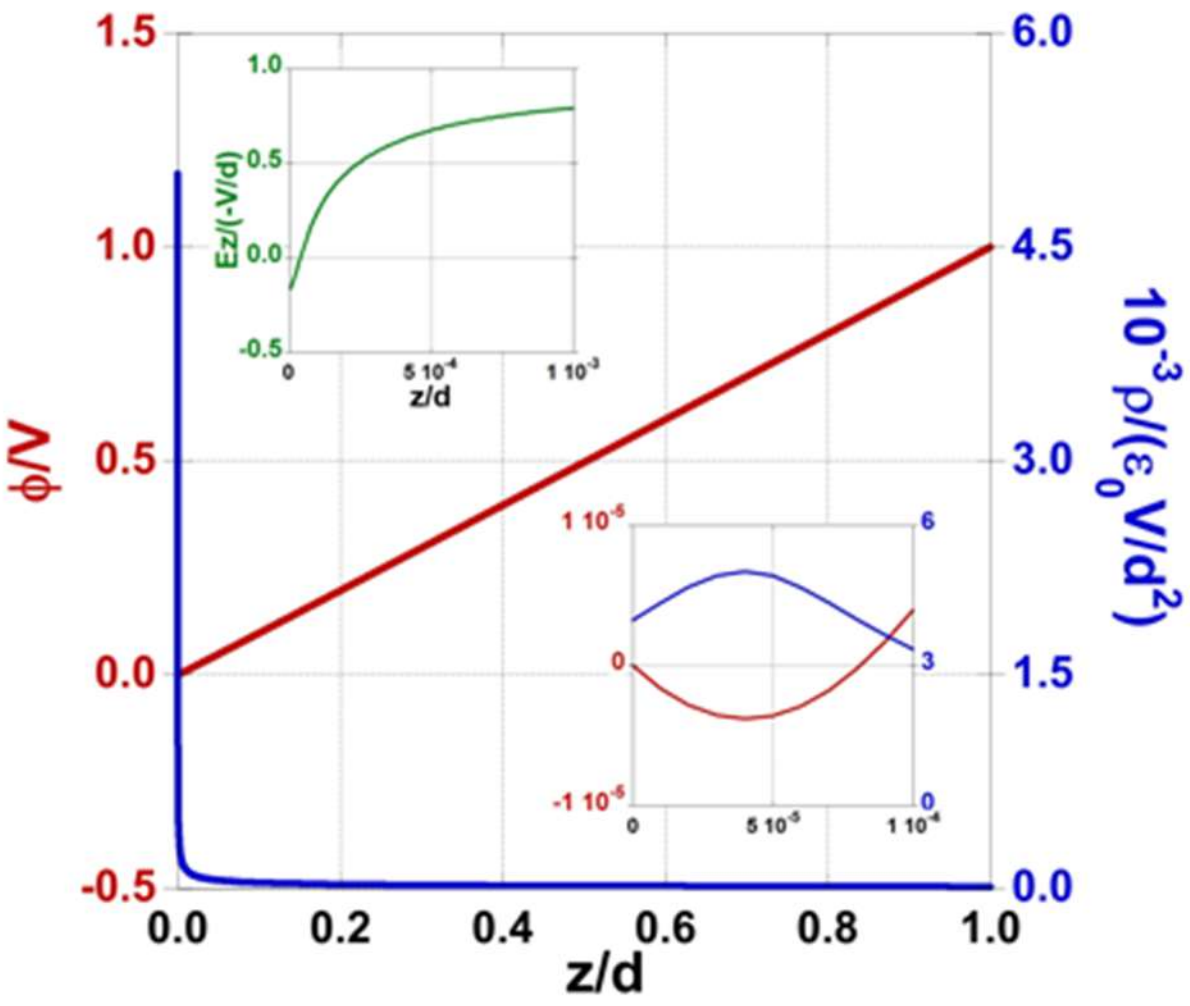

E10
(c)

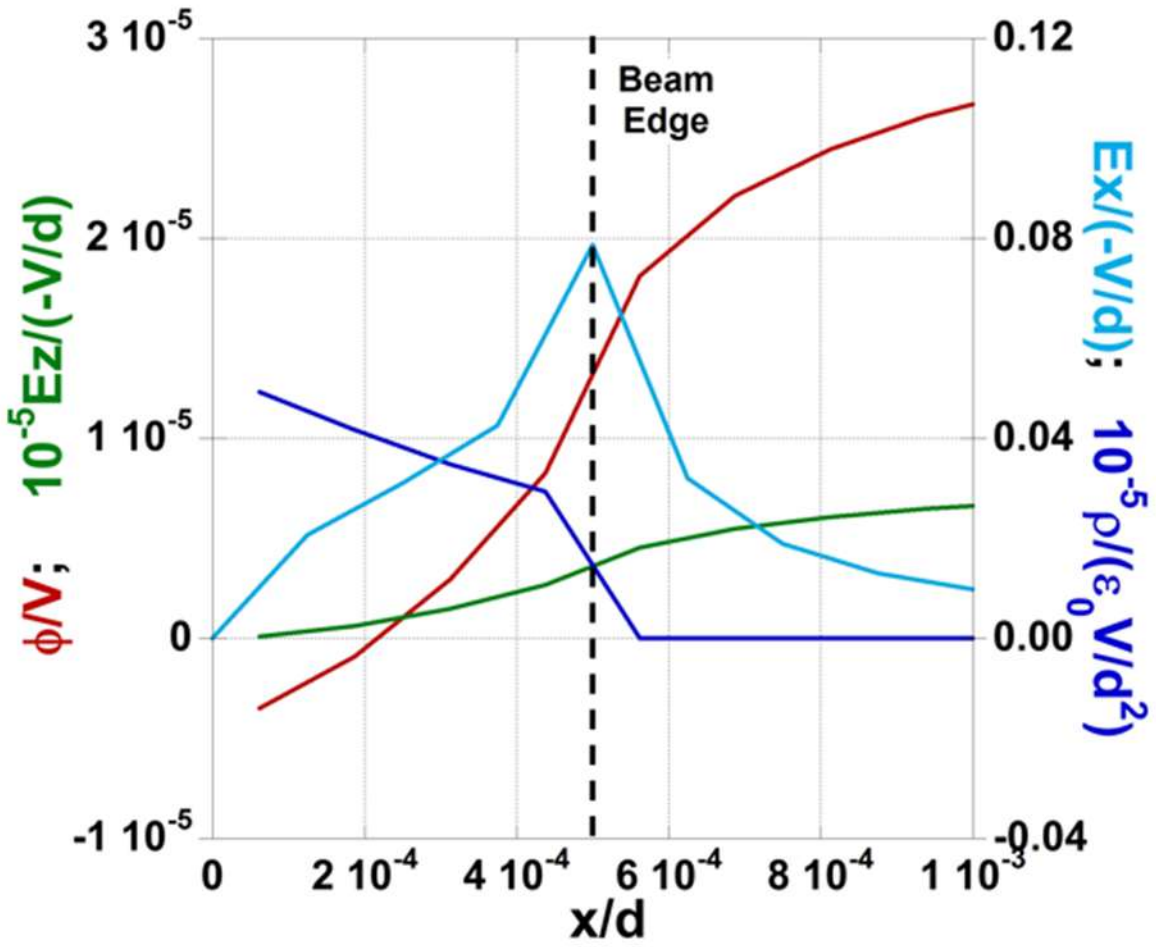

E10
(d)

Fig. 3. Highly resolved POISSON data for Test E7 (a), (b) and for Test E10 (c), (d). For E7, $(\bar{w}, \Delta, \bar{p}) = (10^{-3}, 10^{-2}, 3.2768)$, $(n_x, n_z, d_x, d_z) = (2^{15},\ 2000,\ 10^{-4}d, 5\times10^{-4}d)$, and $\bar{w} \ll 2\pi\Delta$. For E10, $(\bar{w}, \Delta, \bar{p}) = (10^{-3}, 10^{-5}, 2.048)$, $(n_x, n_z, d_x, d_z) = (2^{14},\ 10^{5},\ 1.25\times10^{-4}d, 10^{-5}d)$, and $\bar{w} \gg 2\pi\Delta$. Here, $n_z$ is the number of numerical cells across the diode, $n_x$ is the number of numerical cells from $x = -p/2$ to $x = +p/2$, $d_z = d/n_z$ and $d_x = p/n_x$ are the respective cell size. Plotted in parts (a) and (c) are the normalized potential and charge density as a function of z at x = 0, with their expanded view near the potential minimum, $z = z_{min}$, displayed in the lower inset. The upper inset shows the normalized $E_z$ vs $z$ near the potential minimum which enables an accurate evaluation of $\partial E_z/\partial z$. Plotted in parts (b) and (d) are the normalized potential, charge density, $E_x$, and $E_z$ as a function of $x$, at $z = z_{min}$. They enable an accurate evaluation of $\partial E_x/\partial x$. The POISSON code yields the following normalized values of $(\bar{z}_{min}, \bar{\phi}_{min}, \bar{\rho}, \partial\bar{E}_x/\partial\bar{x}, \partial\bar{E}_z/\partial\bar{z})$ at the potential minimum on the centerline: $(0.0095, -0.00569, 2002, -1905, -97)$ for E7, and $(4\times10^{-5}, -3.8\times10^{-6}, 5014, -165, -4944)$ for E10. The normalized Poisson equation reads, $\partial\bar{E}_x/\partial\bar{x} + \partial\bar{E}_z/\partial\bar{z} = -\bar{\rho}$. Note that $|\partial\bar{E}_x/\partial\bar{x}| \gg |\partial\bar{E}_z/\partial\bar{z}|$ in E7 and $|\partial\bar{E}_x/\partial\bar{x}| \ll |\partial\bar{E}_z/\partial\bar{z}|$ in E10 and therefore these two cases have very different scalings for the 2D limiting current density, as also displayed in Figs. 2(a) and 2(d).

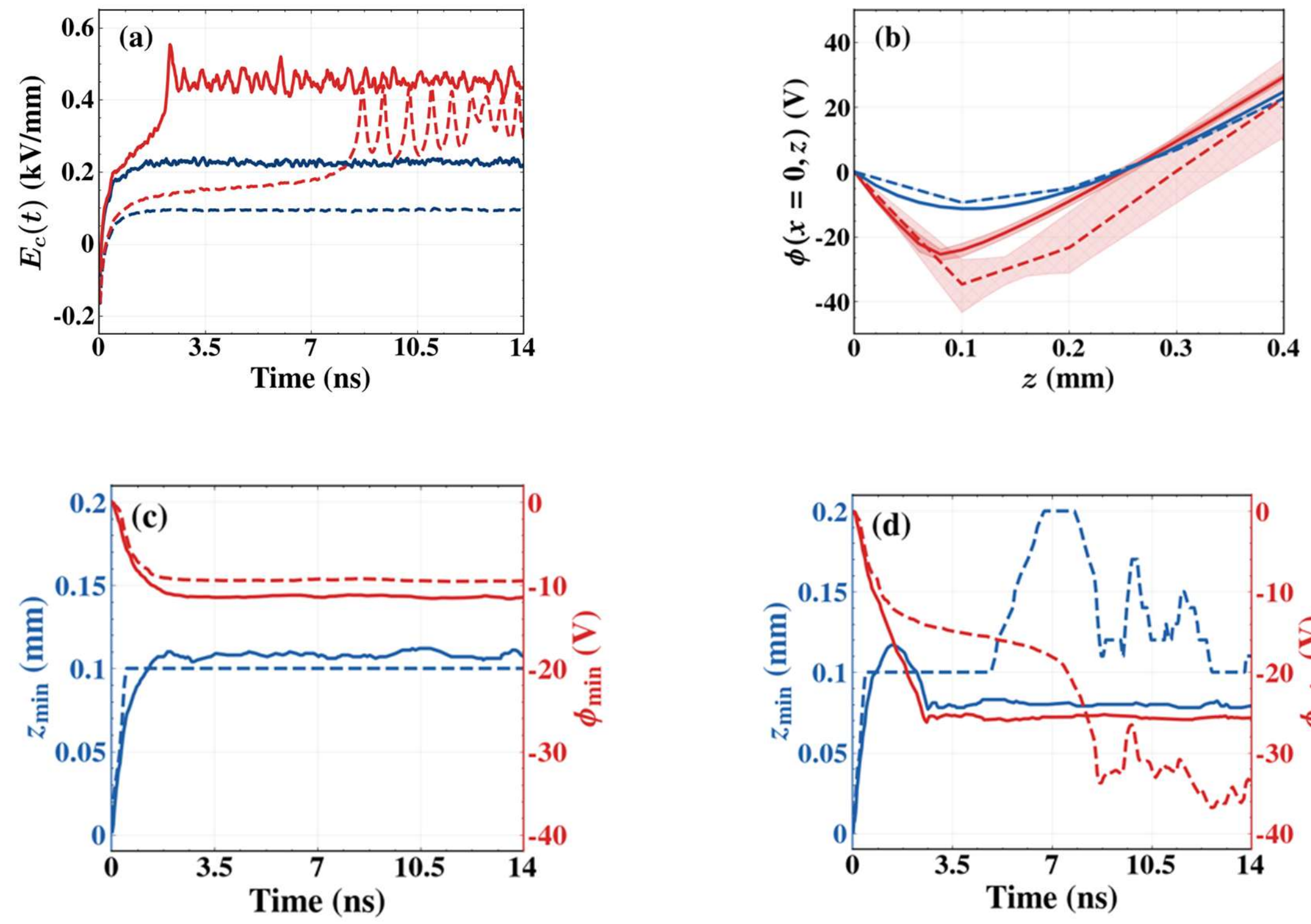

Fig. 4. XOOPIC results for Test E3 $[(\bar{w}, \Delta, \bar{p}) = (0.01, 0.01, 2)]$, showing temporal-spatial evolution in simulations with high and low resolution, at injection current slightly below and above the critical value.

(a). Time evolution of the electric field at the center of the emitting strip, for currents below and above the critical current density $J_{\text{crit}} = 32.75 J_{\text{CL}}$. Blue curves correspond to the subcritical case $J = 32 J_{\text{CL}}$, while red curves correspond to the supercritical case $J = 33.25 J_{\text{CL}}$. Solid lines are for the high-resolution simulation ($n_z = 1000$), and dashed lines represent the low-resolution case ($n_z = 100$). Below $J_{\text{crit}}$, the field relaxes to a stable quasi-steady solution with small fluctuations. As the current approaches and exceeds $J_{\text{crit}}$, the steady solution disappears and the system transitions to a higher-field state with sustained oscillations. Note the abrupt transition at $J_{\text{crit}}$.

(b). Axial potential profile at the center of the emitting strip, $\phi(x = 0,\ z)$, obtained from 20 snapshots taken between $t =$ 10 and 15ns. Solid curves correspond to the high-resolution grid ($n_z = 1000$) and dashed curves to the low-resolution grid ($n_z = 100$). Blue denotes the subcritical case ($J = 32 J_{\text{CL}} < J_{\text{crit}}$), while red denotes the supercritical case ($J = 33.25 J_{\text{CL}} > J_{\text{crit}}$). Shaded regions show the min–max excursion of $\phi(x = 0,\ z)$ over the sampling window. The supercritical case exhibits a substantially deeper potential minimum, while the low-resolution simulation shows larger excursions of the potential profile.

(c). Time evolution of the position of the potential minimum ($z_{\min}$, left axis, blue), and the corresponding minimum potential value ($\phi_{\min}$, right axis, red), for the subcritical case $J = 32 J_{\text{CL}} < J_{\text{crit}}$ . Solid lines denote the high-resolution simulation ($n_z =$ 1000) while dashed lines represent the low-resolution case ($n_z = 100$). This regime shows a relatively stable $z_{\min}$ and $\phi_{\min}$ after the initial transient.

(d). Time evolution of the position of the potential minimum ($z_{\min}$, left axis, blue), and the corresponding minimum potential value ($\phi_{\min}$, right axis, red), for the supercritical case $J = 33.25 J_{\text{CL}} > J_{\text{crit}}$. Solid lines denote the high-resolution simulation ($n_z = 1000$) while dashed lines represent the low-resolution case ($n_z = 100$). The high-resolution solution shows a relatively stable $z_{\min}$ and $\phi_{\min}$ after the initial transient. In contrast, the low-resolution simulation exhibits large excursions in both the position and depth of the potential minimum, especially at later times.

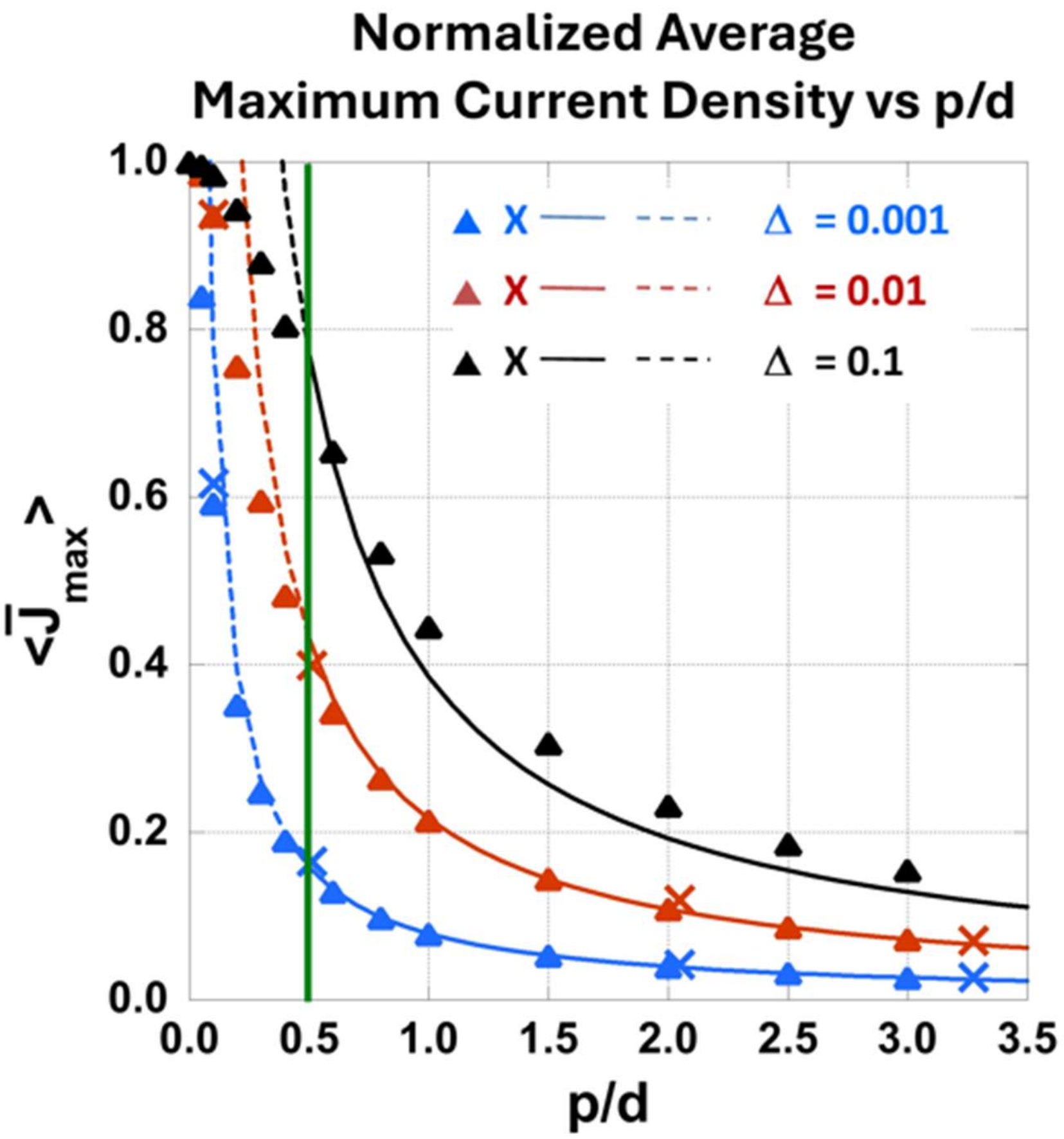

Fig. 5. Plot of $<\bar{J}_{max}> = (\bar{w}/\bar{p})J_{max}/(J_{CL}f_J)$, which is the maximum current density averaged over one period and normalized to the classical 1D Child-Langmuir current density modified by the Jaffe factor, as a function of $\bar{p} = p/d$ for $\Delta$ = 0.001, 0.01, and 0.1, shown respectively in blue, red, and black. The solid and dashed curves plot the scaling law, Eq. (4), which is obtained assuming $\bar{p} \to \infty$. The crosses are the data from the POISSON code, fixing $\bar{w} = 0.001$ so that $\bar{w} \ll 2\pi\Delta$ on all crosses. The triangles are obtained from the integral equation approach assuming $\bar{w} \to 0$ as studied in Chernin et al. [40], whose numerical results (the triangles) are reproduced here. It is remarkable that the scaling law (4), obtained under the assumption $\bar{p} \to \infty$, turns out to be an excellent approximation to the POISSON code results and to the integral equation approach for all $\bar{p} \geq 0.5$, i.e., to the right of the green vertical line in this figure. As $\bar{p} \to 0$, all triangles yield $<\bar{J}_{max}> \to 1$, as mathematically proven in [40]. The parameters used to obtain the POISSON code results represented by the eight (8) crosses are included in Appendix D. Note that the crosses at $(\bar{p}, \Delta) = (2.048, 0.01), (3.277, 0.01), (3.277, 0.001)$ correspond to, respectively, Tests E3, E7, E8. In addition to confirming the remarkable agreements between the different approaches for a wide range of sheet separation, this figure gives the maximum anode current in a 2D periodic array in terms of the packing density, $p/d$. It also illustrates the electrostatic shielding effect (or its lack) by neighboring emitters.

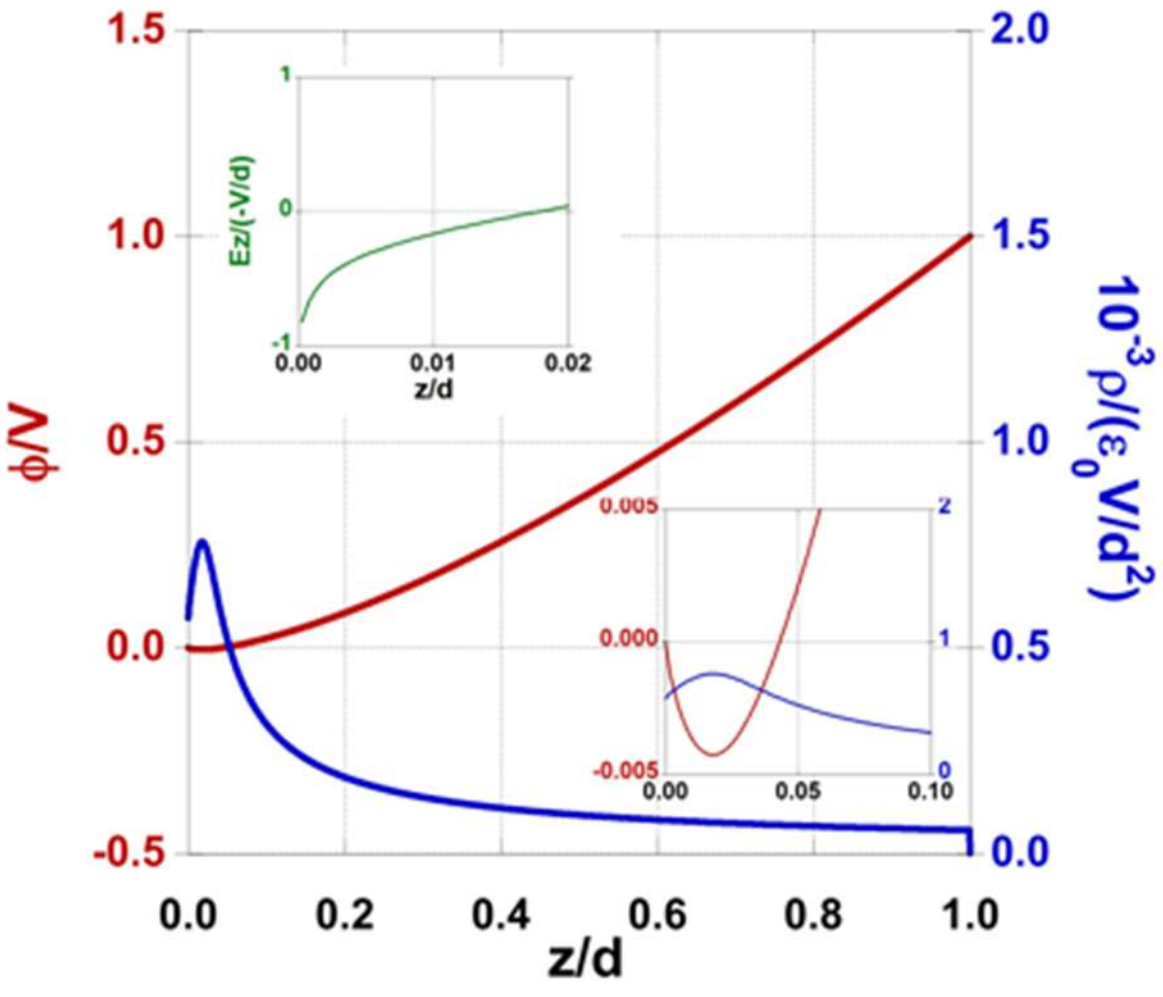

$p/d\ =\ 0.1$

**(a)**

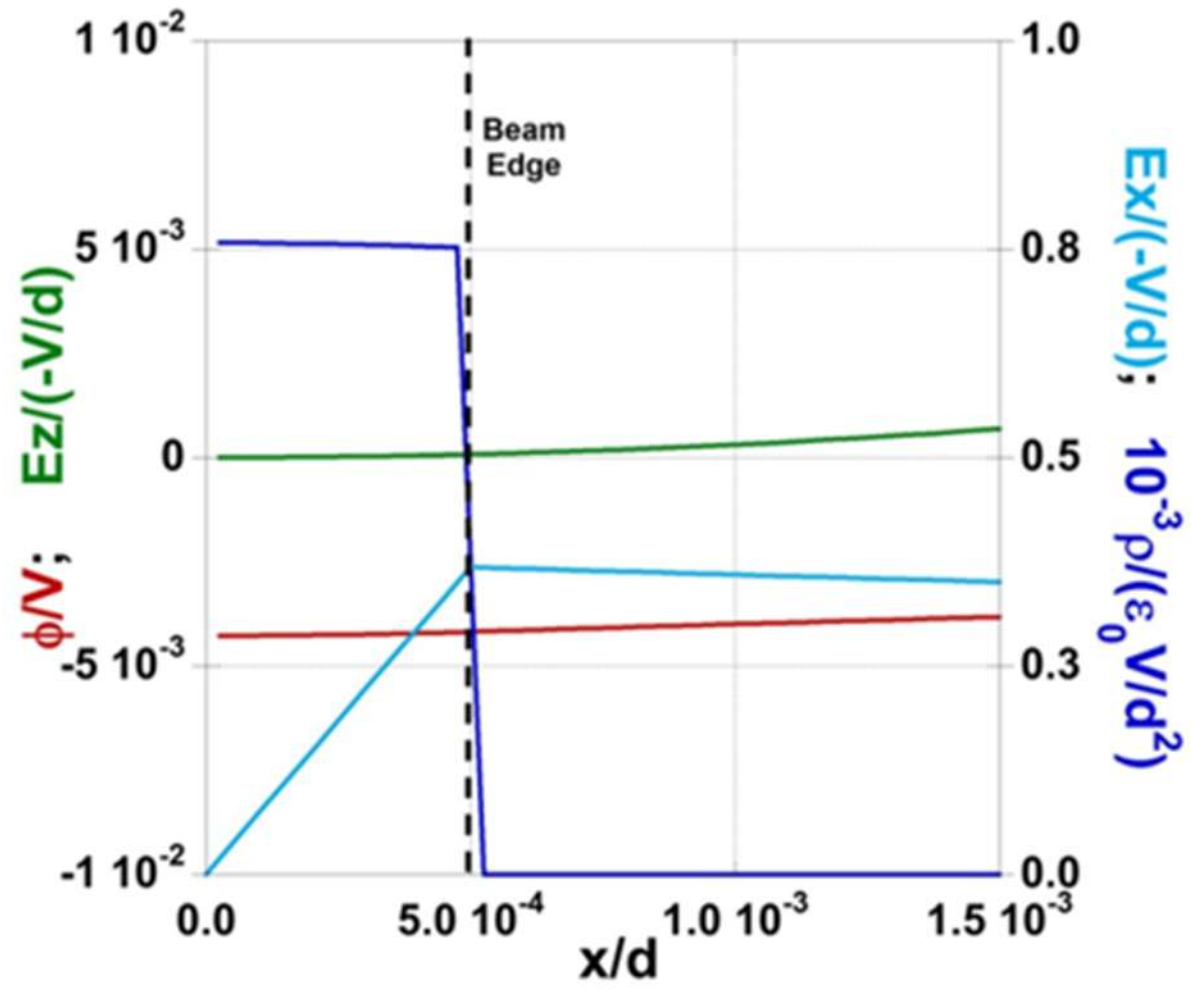

$p/d\ =\ 0.1$

**(b)**

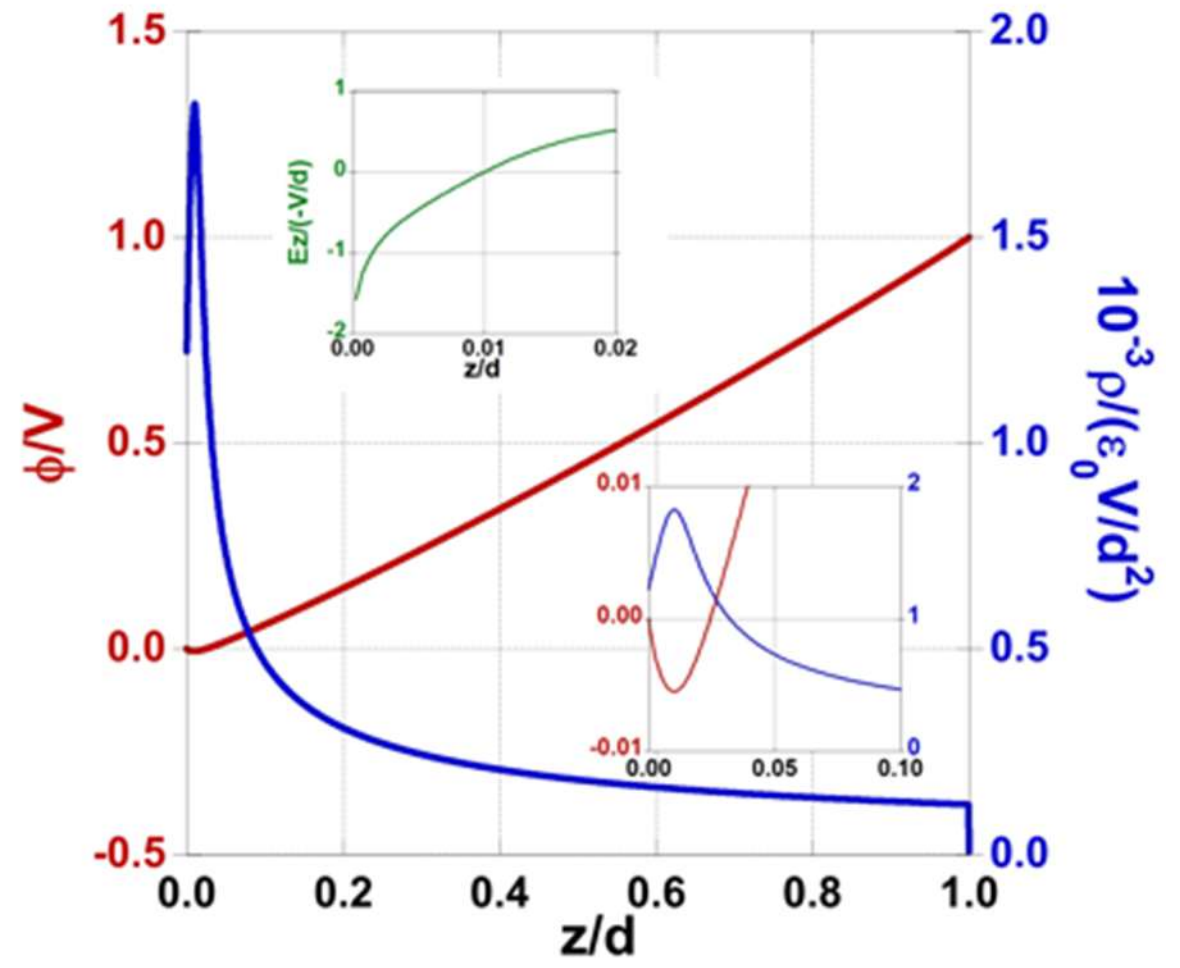

$p/d\ =\ 0.5$

**(c)**

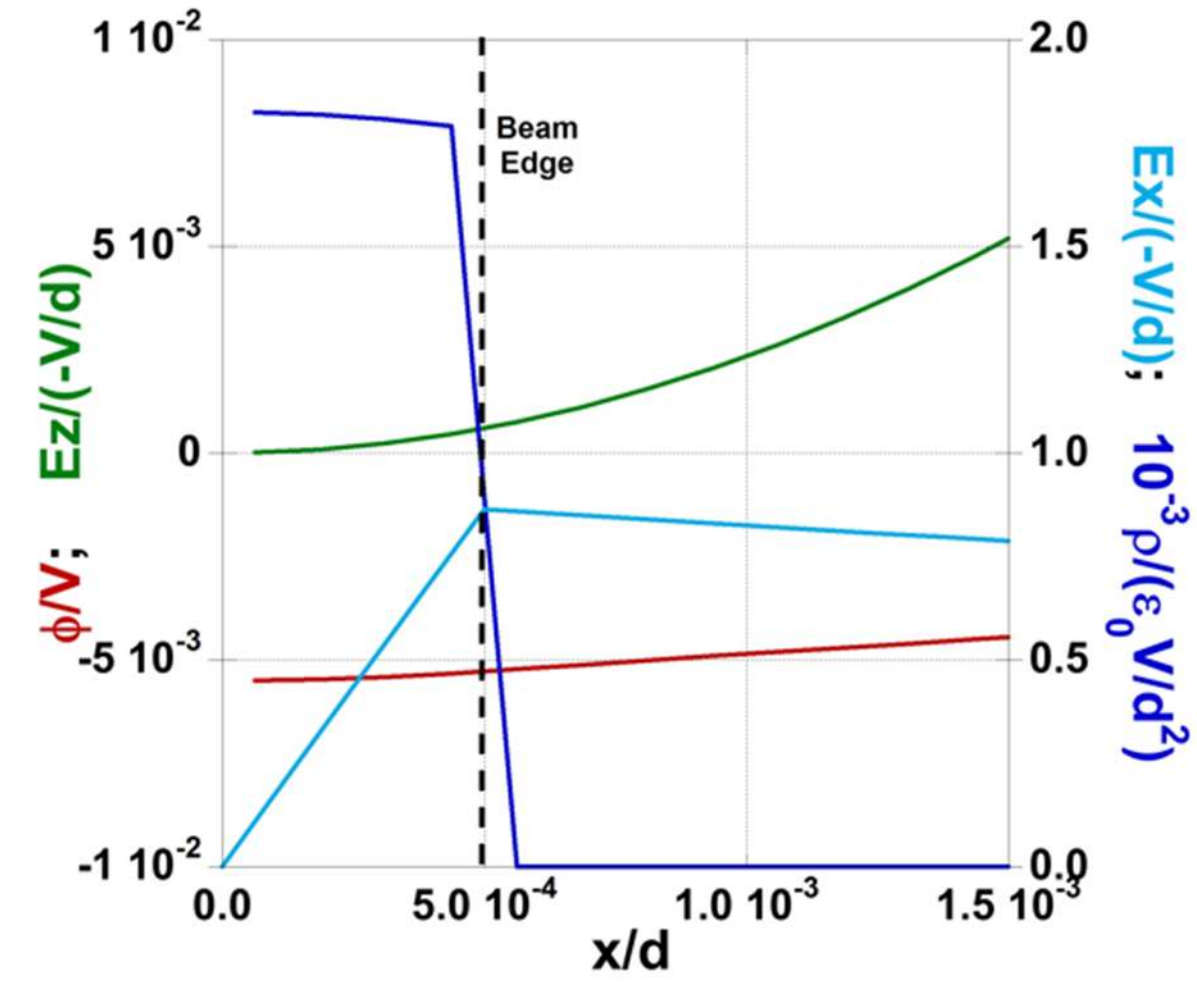

$p/d\ =\ 0.5$

**(d)**

Fig. 6. Highly resolved POISSON data for the regimes of more closely spaced periodic sheets, $p/d \ll 1$ and $p/d < 1$, fixing $\Delta\ =\ 0.01$, and $w/d\ =\ 0.001$ ($\ll 2\pi\Delta$). For Figs. 6a and 6b, we set $p/d\ =\ 0.1$. For Figs. 6c and 6d, we set $p/d\ =\ 0.5$. Plotted in parts (a) and (c) are the normalized potential and charge density as a function of $z$ at $x\ =\ 0$, with their expanded view near the potential minimum, $z =\ z_{min}$, displayed in the lower inset. The upper inset shows the normalized $E_z$ vs $z$ near the potential minimum which enables an accurate evaluation of $\partial E_z/\partial z$. Plotted in parts (b) and (d) are the normalized potential, charge density, $E_x$, and $E_z$ as a function of $x$, at $z =\ z_{min}$. They enable an accurate evaluation of $\partial E_x/\partial x$. Interestingly, the value of $J_{max}/J_{CL}$ for the $p/d\ =\ 0.1$ case behaves like the 1D Child-Langmuir law (i.e., as if the sheets are infinitely closely-spaced) to within ~ 10 percent, as shown by the cross at $(\bar{p}, \Delta) = (0.1, 0.01)$ in Fig. 5. In contrast, the value of $J_{max}/J_{CL}$ for the $p/d\ =\ 0.5$ case behaves like the 2D Child-Langmuir law (i.e., as if each sheet is treated as infinitely far from each other), also to within 10 percent, as shown by the cross at $(\bar{p}, \Delta) = (0.5, 0.01)$ in Fig. 5 in comparison with the red solid curve. This is an unexpected, surprising finding. Despite the marked differences in the dependence of $J_{max}/J_{CL}$ in 1D ($p/d\ =\ 0.1$) vs 2D ($p/d\ =\ 0.5$) descriptions, the field patterns between Figs. 6a and 6c, and between Figs. 6b and 6d, are qualitatively similar because these two cases both belong to the regime $w/d\ =\ 0.001 \ll 2\pi\Delta =\ 0.0628$. The POISSON parameters used for these two cases ($p/d\ =\ 0.1$ and 0.5) are listed in Appendix D.

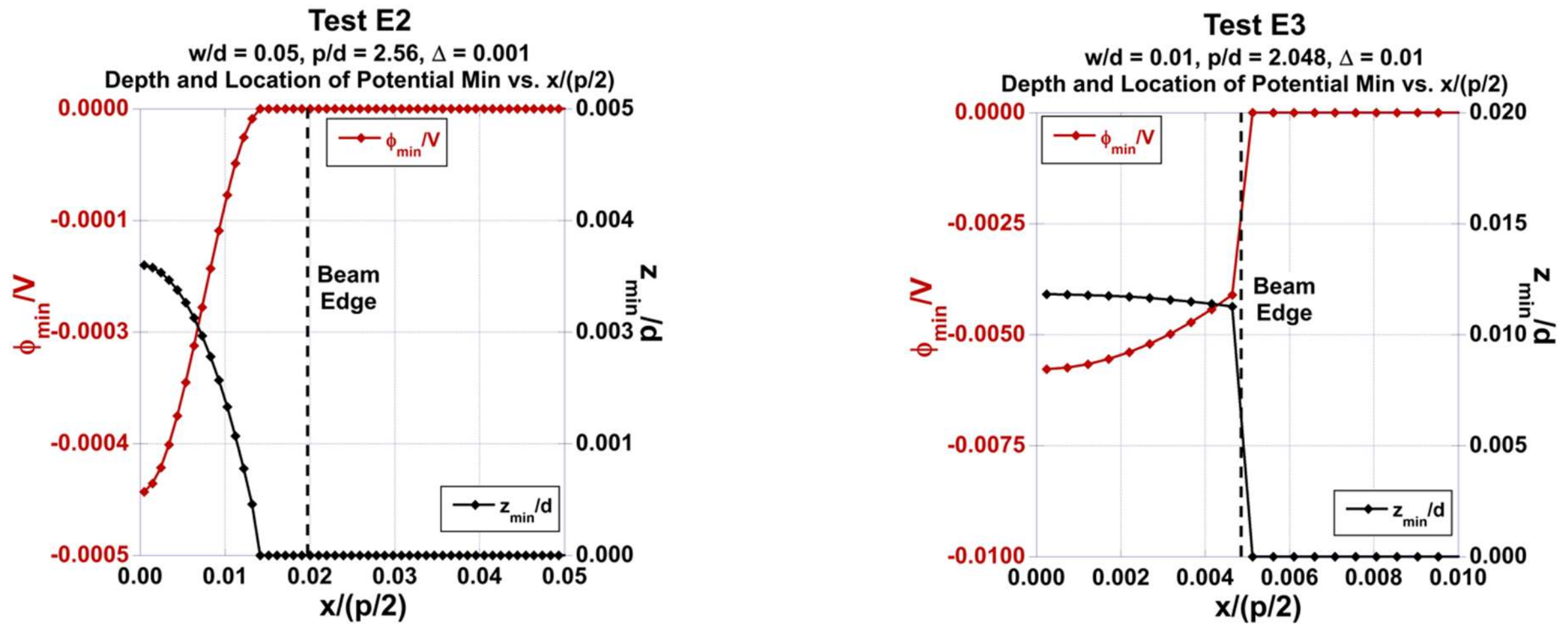

**(a)** **(b)**

Fig. 7. The location and the value of the potential minimum as a function of $x$ (along the cathode surface, obtained from the POISSON code) at the critical current density for (a) Test E2, and (b) Test E3. Note that Test E2 is in the regime $w/d \gg 2\pi\Delta$ and that Test E3 is in the regime $w/d \ll 2\pi\Delta$ (see also Figs. 2a,b). By resolving the potential minimum for each value of $x$, we show the huge contrasts exhibited in the two regimes, (a) and (b).